\documentclass[12pt,a4paper, table, xcdraw]{article}
\pdfoutput=1
\usepackage{geometry}
\UseRawInputEncoding
\usepackage{amsmath}
\usepackage{amsmath}
\usepackage{amssymb}
\usepackage[dvips]{graphicx}
\usepackage{cite}
\usepackage{pstricks}
\usepackage{bm}
\usepackage{bbm}
\usepackage{pbox}
\usepackage{placeins}
\usepackage{graphicx}
\usepackage{caption}
\usepackage{subcaption}
\usepackage[T1]{fontenc}
\usepackage{footnote}
\usepackage{pdfpages}
\usepackage{hhline}
\usepackage{multirow}
\usepackage{multicol}
\usepackage{enumitem}
\usepackage{multirow}
\usepackage{amsmath}
\usepackage{relsize}
 \usepackage{graphicx}
\usepackage[toc,page]{appendix}
\usepackage[english]{babel}
\allowdisplaybreaks
\usepackage{xcolor}
\usepackage{float}
\usepackage{slashed}
\usepackage{cases}
\usepackage{empheq}

\definecolor{MyDarkBlue}{rgb}{0.1, 0.3, 0.8} 
\definecolor{SBlue}{rgb}{0.2, 0.4, 0.8} 
\definecolor{MyLightBlue}{rgb}{0.22,0.51,0.99}
\definecolor{MyGreen}{rgb}{0.0, 0.5, 0.3}
\definecolor{BrickRed}{rgb}{0.8, 0.25, 0.33}
\usepackage[colorlinks=true,linkcolor=blue,citecolor=MyDarkBlue,
urlcolor=BrickRed,bookmarksnumbered=true,bookmarksopen]{hyperref}
\hypersetup{colorlinks, citecolor=SBlue,linkcolor=MyGreen, urlcolor=MyDarkBlue}
\begin{document}
\vspace*{-0.2in}
\begin{flushright}
\end{flushright}
\begin{center}
{\Large \bf
Probing Neutrino Mass Models  through Resonances \\[0.06in] at  Neutrino Telescopes
}
\end{center}
\renewcommand{\thefootnote}{\fnsymbol{footnote}}
\begin{center}
{
{}~\textbf{K.S. Babu$^a$}\footnote{ E-mail: \textcolor{MyDarkBlue}{babu@okstate.edu}},
{}~\textbf{P.S. Bhupal Dev$^b$}\footnote{ E-mail: \textcolor{MyDarkBlue}{bdev@wustl.edu}},
{}~\textbf{Sudip Jana$^c$}\footnote{ E-mail: \textcolor{MyDarkBlue}{sudip.jana@mpi-hd.mpg.de}}
}
\vspace{0.3cm}
{
\\\em $^a$Department of Physics, Oklahoma State University, Stillwater, OK 74078, USA
\\
$^b$Department of Physics and McDonnell Center for the Space Sciences, \\ Washington University, St. Louis, MO 63130, USA
\\
$^c$Max-Planck-Institut f{\"u}r Kernphysik, Saupfercheckweg 1, 69117 Heidelberg, Germany
} 
\end{center}
\renewcommand{\thefootnote}{\arabic{footnote}}
\setcounter{footnote}{0}
\thispagestyle{empty}

\begin{abstract}
 We study the detection prospects of relatively light charged scalars in radiative Majorana neutrino mass models, such as the Zee model and its variants using scalar leptoquarks, at current and future neutrino telescopes. In particular, we show that these scalar mediators can give rise to Glashow-like resonance features in the ultra-high energy neutrino (UHE) event spectrum at the IceCube neutrino observatory and its high-energy upgrade IceCube-Gen2. The same scalars can also give rise to  observable non-standard neutrino interactions (NSI), and we show that 
the UHE neutrinos provide a complementary probe of NSI. We also discuss an interesting possibility of producing such resonances by incoming sterile neutrino components in the case where neutrinos are pseudo-Dirac particles. 
\end{abstract}
\newpage
\setcounter{footnote}{0}
\section{Introduction}
Understanding the origin of small but nonzero neutrino masses required to explain the observed neutrino oscillation data~\cite{ParticleDataGroup:2020ssz} is of fundamental importance in elementary particle physics. Within the Standard Model (SM) neutrinos are precisely massless to all orders in perturbation theory owing to the gauge structure and its particle content; consequently, one must go beyond the SM to explain neutrino masses. 

One obvious way to introduce  neutrino mass is by adding their right-handed partners (the so-called `sterile' neutrinos) to the theory and by utilizing the same Higgs mechanism as for the charged fermions in the SM. Neutrinos would then be Dirac fermions and lepton number would remain exact in this case. However, the fact that the neutrino masses need to be at the sub-eV scale requires the corresponding Dirac Yukawa couplings to be $Y_\nu \lesssim 10^{-12}$. Although there is nothing wrong with this picture, there is no satisfactory explanation of such tiny Yukawa couplings. Moreover, this is a rather boring scenario from the phenomenological perspective. 

On the other hand, if neutrinos are Majorana particles, their masses could arise from effective higher-dimensional lepton number violating (LNV) operators. This is the case with the seesaw mechanism, induced by the dimension-5  operator~\cite{Weinberg:1979sa} 
\begin{equation}
    {\cal O}_1 \ = \ L^iL^j H^kH^l \epsilon_{ik} \epsilon_{jl} \, ,
    \label{O1}
\end{equation}
where $L$ stands for the lepton doublet, and $H$ for the Higgs doublet, with $i,j,k,l$ denoting $SU(2)_L$ indices, and $\epsilon_{ik}$ is the $SU(2)_L$ antisymmetric tensor.  In this case the inverse mass dimension that multiplies this operator is of order $M^{-1} \sim (10^{15}~{\rm GeV})^{-1}$, which would make it difficult to directly probe this mechanism.  Another alternative for generating small neutrino masses is by quantum corrections~\cite{Zee:1980ai, Hall:1983id, Zee:1985id, Babu:1988ki, Krauss:2002px, Cai:2014kra, Babu:2019mfe} (for a review, see Ref.~\cite{Cai:2017jrq}).  In these radiative neutrino mass models, the tree-level Lagrangian does not generate ${\cal O}_1$ owing to the particle content or symmetries present in the model.  If such a model has lepton number violation, then small Majorana masses for neutrinos may arise at one-loop, two-loop, or higher loop level, depending on the model details, which will have an appropriate loop suppression factor, and typically a chiral suppression factor involving a light fermion mass as well. Therefore, to account for the observed neutrino mass spectrum, these models typically require (sub-)TeV scale new physics (unlike the simple seesaw models) that can be directly probed in laboratory experiments, enabling direct tests of the origin of neutrino mass. In this review, we will focus on some phenomenological aspects of the mediators of radiative neutrino mass generation, including their unique signatures at neutrino telescopes using ultra-high energy (UHE) neutrinos~\cite{Babu:2019vff}. In addition, we demonstrate that these new scalar resonances provide a new probe of neutrino non-standard interactions (NSI)~\cite{Wolfenstein:1977ue}, complementary to other laboratory probes of NSI.   

The observation of UHE neutrinos  at IceCube~\cite{Aartsen:2013bka, Aartsen:2013jdh, Aartsen:2014gkd} has begun a new era in high-energy neutrino astrophysics. Understanding all aspects of these UHE neutrino events is extremely important for both astrophysics and partcile physics communities~\cite{Anchordoqui:2013dnh, Ahlers:2018mkf}. This includes identifying their sources, energy flux, flavor composition, propagation, and detection. An isotropic, single power-law astrophysical neutrino flux parametrized as
\begin{align}
    \frac{d\Phi_\nu}{dE} \ = \ \Phi_{\rm astro}\left(\frac{E_\nu}{100~{\rm TeV}}\right)^{-\gamma_{\rm astro}}\cdot 10^{-18}~{\rm GeV}^{-1}{\rm cm}^{-2}{\rm s}^{-1}{\rm sr}^{-1}
    \label{eq:flux}
\end{align}
provides a good description of the high-energy starting event (HESE) component of the IceCube data, with the latest 7.5-year best-fit values of $\Phi_{\rm astro} =6.37^{+1.46}_{-1.62}$ and $\gamma_{\rm astro}=2.87^{+0.20}_{-0.19}$ at $68.3\%$ confidence level (CL)~\cite{IceCube:2020wum}. With more data expected soon and with multimessenger probes of the astrophysical sources, it is likely that the precision on the UHE neutrino spectrum measurement will significantly improve in the not-so-distant future, more so with the planned IceCube-Gen2 upgrade~\cite{IceCube-Gen2:2020qha}. 
Therefore, any anomalous features in the observed UHE neutrino event spectrum could  be used as a probe of fundamental physics.  

Mediators of radiative neutrino mass mechanism could provide one such anomalous feature in the form of a new resonance.  The purpose of this review is to show that such a new resonance can arise naturally in a broad class of radiative neutrino mass models and to study their signatures at IceCube and other UHE neutrino detectors. While we focus primarily on the Zee model~\cite{Zee:1980ai} of radiative neutrino mass and its leptoquark variants~\cite{Cai:2014kra, Babu:2019mfe}, our analysis is more general and is applicable to a variety of radiative neutrino mass models which could potentially lead to observable signals at IceCube. This includes one-loop~\cite{Zee:1980ai, Cai:2014kra}, two-loop~\cite{Zee:1985id, Babu:1988ki} and three-loop~\cite{Krauss:2002px, Aoki:2008av, Gustafsson:2012vj} radiative models with color neutral scalars, as well as one-loop~\cite{AristizabalSierra:2007nf,Cai:2014kra, Cheung:2016fjo, Dorsner:2017wwn,  Popov:2019tyc,Bigaran:2019bqv,Babu:2020hun,Chang:2021axw, Zhang:2021dgl}, two-loop~\cite{Babu:2010vp, Babu:2011vb, Kohda:2012sr, Angel:2013hla}, and three-loop~\cite{Nomura:2016ezz} radiative neutrino mass models utilizing leptoquarks with or without color neutral scalars. In all these models it is almost always necessary to have new scalar bosons for neutrino mass generation using effective higher-dimensional operators~\cite{Babu:2001ex}. Another general feature of all these models is that the light neutrino mass matrix is proportional to a product of two Yukawa coupling matrices multiplying certain charged fermion mass matrix.  In our analysis we shall assume that one of the Yukawa coupling matrices has entries of order one, which could  be made consistent with small neutrino masses by choosing the other Yukawa coupling matrix appropriately small.

The rest of the review is organized as follows: In Section~\ref{sec:mediator}, we discuss various possibilities for a new scalar resonance at IceCube in the context of radiative neutrino mass models. In Section~\ref{sec:zee}, we consider the prototypical example of the Zee scalar, while in Section~\ref{sec:LQ} we consider the scalar leptoquarks. In Section~\ref{sec:pseudo}, we briefly discuss resonances associated with pseudo-Dirac neutrino scenario. Finally, we conclude in Section~\ref{sec:con}. 
\section{New Scalar Resonances at IceCube} \label{sec:mediator}
The only SM resonance IceCube is truly sensitive to is the Glashow resonance~\cite{Glashow:1960zz}, where electron anti-neutrinos hitting the target electrons in ice produce an on-shell $W$-boson: $\bar{\nu}_e e^-\to W^-\to {\rm anything}$. The required energy of the incoming antineutrino for this resonance to happen is $E_\nu=m_W^2/2m_e= 6.3$ PeV. Only recently, IceCube has reported the detection of a particle shower consistent with being created at the Glashow resonance, with a deposited shower energy of $6.05 \pm 0.72$ PeV~\cite{IceCube:2021rpz}. On the other hand, the possibility of detecting an analogous $Z$-boson resonance (the so-called $Z$-burst) at IceCube due to UHE (anti-)neutrinos interacting with non-relativistic relic neutrinos~\cite{Weiler:1982qy} is rather bleak, as the required incoming neutrino energy in this case turns out to be $E_\nu=m_Z^2/2m_\nu \gtrsim 10^{23}$ eV (corresponding to $m_\nu\lesssim 0.05$ eV), which is well beyond the Greisen$-$Zatsepin$-$Kuzmin (GZK) cut-off energy of $\sim 5\times 10^{19}$ eV for the progenitor UHE cosmic rays~\cite{Greisen:1966jv, Zatsepin:1966jv}. Similarly, lower-energy charged vector meson resonances~\cite{Paschos:2002sj} like $\bar{\nu}_ee^-\to \rho^-\to \pi^-\pi^0$ will most likely remain out-of-reach of IceCube (and its future upgrades), primarily because of the large deep-inelastic-scattering (DIS) background~\cite{Brdar:2021hpy}. The neutral vector meson resonances like $\bar{\nu}_\ell\nu_\ell \to \rho^0\to \pi^+\pi^-$ are further suppressed by the relic neutrino density  (compared to the typical electron density in matter)~\cite{Dev:2021tlo}.

Other interesting possibilities arise in beyond SM scenarios involving secret neutrino interactions with a light (MeV-scale) $Z'$~\cite{Araki:2014ona, Araki:2015mya, Kamada:2015era, DiFranzo:2015qea} or light neutrinophilic neutral scalar~\cite{Ioka:2014kca, Ng:2014pca, Ibe:2014pja}, in which case the resonance could again fall in the multi-TeV to PeV range which will be accessible at IceCube. Yet another alternative is the possibility of heavier (100 GeV to TeV-scale) resonances induced by the scalar mediators in radiative neutrino mass models, which is the focus of this review. These mediators could be either charged color-singlets such as the Zee scalars~\cite{Babu:2019vff} inducing neutrino-electron interactions,  or charged colored particles such as leptoquarks~\cite{Barger:2013pla, Dey:2017ede, Becirevic:2018uab, Dorsner:2019vgp, Huang:2021mki} and squarks in $R$-parity violating supersymmetry~\cite{Carena:1998gd,Dev:2016uxj, Collins:2018jpg, Dev:2019ekc} inducing neutrino-nucleon interactions.

We follow the nomenclature proposed in Ref.~\cite{Babu:2019mfe} and only consider those radiative neutrino mass models containing at least one SM particle inside the loop diagram generating neutrino mass (the so-called type-I radiative models). These models can be described by effective higher-dimensional LNV  operators, similar to Eq.~\eqref{O1}.  A prototypical example is the Zee model \cite{Zee:1980ai} which introduces a second Higgs doublet $\Phi({\bf 1},{\bf 2},1/2)$ and a charged $SU(2)_L$-singlet scalar $\eta^+({\bf 1},{\bf 1},1)$ to the SM (with the $SU(3)_c\times SU(2)_L\times U(1)_Y$ charges shown in parentheses). This leads to a one-loop neutrino mass via the effective LNV $(\Delta L = 2)$ dimension-7 operator
\begin{equation}
    {\cal O}_2 \ = \ L^i L^j L^k e^c H^l \epsilon_{ij} \epsilon_{kl} \, ,
    \label{O2}
\end{equation}
with indices $i,j,..$ referring to $SU(2)_L$, and $e^c$ standing for the $SU(2)_L$ singlet left-handed positron state. Variants of the Zee model can be constructed by replacing the new scalars with colored leptoquark scalars~\cite{Cai:2014kra, Babu:2019mfe}, with one natural realization being $R$-parity breaking supersymmetry~\cite{Hall:1983id}.

\begin{figure}[t!]
\centering
$$
\includegraphics[width=0.4\textwidth]{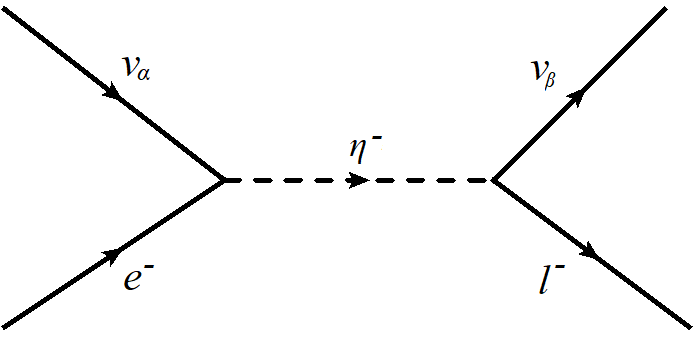} \hspace{0.5in}
\includegraphics[width=0.4\textwidth]{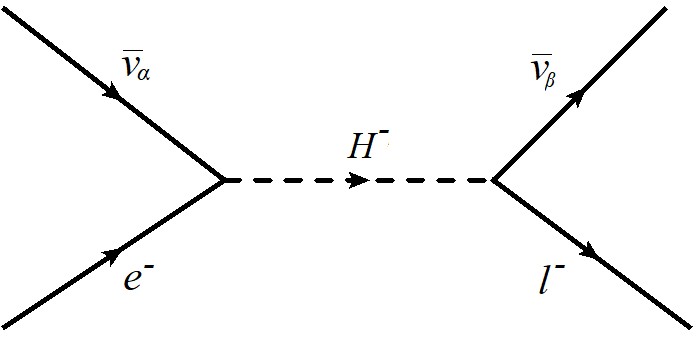}
$$
$$
\includegraphics[width=0.4\textwidth]{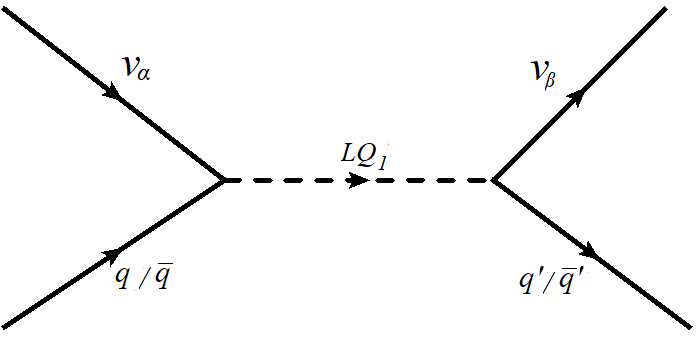} \hspace{0.5in}
\includegraphics[width=0.4\textwidth]{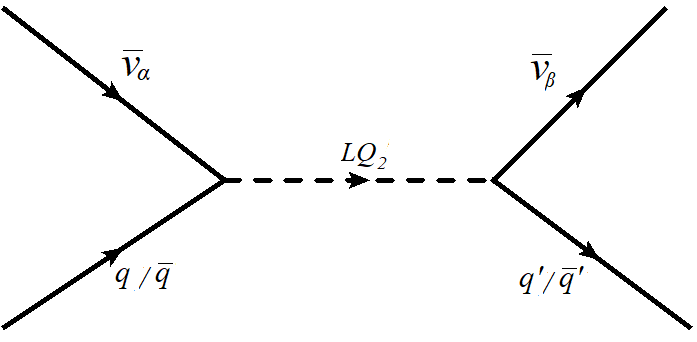}
$$
\caption{Feynman diagrams for resonances at neutrino telescopes: in the top row, we have color-singlet $SU(2)_L$ singlet (left) and doublet scalars, whereas in the bottom row, we have a color-triplet $SU(2)_L$ singlet (left) and doublet (right) scalar leptoquarks. \label{fig:feyn_spectra}
}
\end{figure}

In all these models there are new scalar mediators, which also lead to unique resonance features in UHE neutrino interactions with matter (either up- or down-quarks,  or electrons). For future reference, we show them in Fig.~\ref{fig:feyn_spectra} and also list them in Table~\ref{tab:glossory}, along with their field components and corresponding Lagrangian terms responsible for NSI. For a singly-charged scalar, $\eta^+$ and $h^+$ are used interchangeably, to be consistent with the literature. Similarly, we use the standard notation of $S_1$, $S_3$, $R_2$ and $\widetilde{R}_2$ for the scalar leptoquarks to avoid confusion, but for their field components under $SU(2)_L$, we still use the notation from Ref.~\cite{Babu:2019mfe}, where these leptoquark fields were denoted as $\chi^\star$, $\bar{\rho}$, $\delta$ and $\Omega$, respectively. 

As shown in Fig.~\ref{fig:feyn_spectra}, the resonance will be an $SU(2)_L$ singlet for incoming neutrinos (left panels) and  $SU(2)_L$-doublet for an incoming antineutrino (right panels). For (anti)neutrino-electron scattering, the resonance is a color-singlet charged scalar (top row) and for (anti)neutrino-quark scattering, the resonance is a color-triplet scalar leptoquark (bottom row). Depending on the charge of the leptoquark component involved in the resonance, we can have either quark or antiquark in the initial state. Note that the antiquark parton distribution functions (PDF) coming from the sea-quark population can be significant and even dominant over the valence quark contributions at higher energies~\cite{Gandhi:1995tf}. 

Note that Table~\ref{tab:glossory} is not an exhaustive list of mediators for radiative neutrino mass models, because we have omitted the ones not relevant for our discussion of the IceCube signals. For instance, a triplet scalar $\Delta({\bf 1},{\bf 3},{\bf 1})$ can also induce a loop-level neutrino mass via the $L_\alpha L_\beta \Delta$ term, in addition to a tree-level neutrino mass via the type-II seesaw~\cite{Schechter:1980gr,Cheng:1980qt,Mohapatra:1980yp,Lazarides:1980nt}. The singly-charged component of $\Delta$ can in principle be resonantly produced at IceCube via neutrino-electron interactions. However, to see any observable effects, it has to be relatively light $\lesssim$ 100 GeV~\cite{Babu:2019vff}. On the other hand, the electroweak $T$-parameter constraint requires that the mass splitting between the singly-charged and doubly-charged components of $\Delta$ cannot exceed 50 GeV or so~\cite{Kanemura:2012rs, Chun:2012jw}, which implies that to get an IceCube signal for the singly-charged component would also require a relatively light doubly-charged component coupling to electrons, which is however excluded by LHC searches~\cite{ATLAS:2017xqs}. 

Similarly, we do not include vector leptoquarks in our discussion, because there are no compelling UV complete models of radiative neutrino mass with vector leptoquarks. 
In a UV-complete model for vector leptoquarks, such as Pati-Salam or SO(10), which is necessary to make sense of the loop calculations, there are already tree-level contributions to the neutrino mass, so the loop contributions are typically sub-dominant~\cite{Babu:2001ex}.

\begin{table}[!t]
\centering
   { \begin{tabular}{|c|c|}
    \hline \hline
      \textbf{  Particle Content}  & \textbf{Lagrangian term} \\ \hline \hline
   $\eta^+({\bf 1},{\bf 1},1)$ or $h^+({\bf 1},{\bf 1},1)$ & $f_{\alpha\beta}L_\alpha L_\beta\, \eta^+$  or $f_{\alpha\beta}L_\alpha L_\beta\, h^+$\\ \hline 
   $H\left({\bf 1},{\bf 2},\frac{1}{2}\right) = \left(H^+,H^0\right)$ & $Y_{\alpha\beta}L_\alpha \ell^c_\beta \widetilde{H}$ \\ \hline
   $S_1\left(\bar{{\bf 3}},{\bf 1},\frac{1}{3}\right)$ & $\lambda_{\alpha\beta}L_\alpha Q_\beta S_1$ \\ \hline
   $S_3\left(\bar{\bf 3},{\bf 3},\frac{1}{3}\right)=\left({\rho}^{4/3},{\rho}^{1/3},{\rho}^{-2/3}\right)$ & $\lambda'_{\alpha\beta}L_\alpha Q_\beta S_3$ \\ \hline
   $R_2\left({\bf 3},{\bf 2},\frac{7}{6}\right)=\left(\delta^{5/3},\delta^{2/3}\right)$ & $\lambda''_{\alpha\beta}L_\alpha u^c_\beta R_2$ \\ \hline
    $\widetilde{R}_2\left({\bf 3},{\bf 2},\frac{1}{6}\right) =\begin{pmatrix} \omega^{2/3} , \omega^{-1/3} \end{pmatrix}$
   & $\lambda'''_{\alpha\beta}L_\alpha d^c_\beta \widetilde{R}_2$ \\ \hline
   \hline
    \end{tabular}}
     \caption{Summary of new particles, their $SU(3)_c\times SU(2)_L\times U(1)_Y$ quantum numbers (with the non-Abelian charges in boldface), field components and electric charges (in superscript), and corresponding Lagrangian terms responsible for NSI in various type-I radiative neutrino mass models considered here. $\widetilde{H}=i\tau_2 H^\star$ (with $\tau_2$ being the second Pauli matrix.}
     \label{tab:glossory}
    \end{table}
\section{Zee-burst at IceCube} \label{sec:zee}
As our first prototypical example, we consider the Zee model~\cite{Zee:1980ai} -- one of the simplest and most popular radiative neutrino mass models. It contains an $SU(2)_L$-singlet charged scalar field $\eta^\pm$ and an $SU(2)_L$-doublet scalar field $H_2$, in addition to the SM-like Higgs doublet $H_1$. The original version of the Zee model~\cite{Zee:1980ai} has sufficient flexibility so that it is fully consistent with the observed neutrino oscillation data~\cite{Herrero-Garcia:2017xdu, Babu:2019mfe}. It may be noted that the Wolfenstein version of the model~\cite{Wolfenstein:1980sy} which assumes a $Z_2$ symmetry that prevents the leptons from coupling to both $H_1$ and $H_2$ is excluded by oscillation data~\cite{Koide:2001xy, He:2003ih}, since in this case all the diagonal entries of the neutrino mass matrix vanish (in a basis where the charged lepton mass matrix is diagonal). It was shown in Ref.~\cite{Babu:2019mfe} that both of the charged scalars of the model, arising from the singlet $\eta^\pm$ and the doublet $H_2^\pm$ can be as light as ${\cal O}(100~{\rm  GeV})$, while satisfying all theoretical and current experimental constraints. It has also been noted~\cite{Babu:2019mfe} that such light charged scalars may result in a significant diagonal NSI of neutrinos and electrons that can be probed in accelerator and atmospheric neutrino experiments. Thus, the possibility of having a resonance feature with these light charged scalars of the Zee model -- dubbed as `Zee-burst'~\cite{Babu:2019vff} --  provides a new probe of NSI with UHE neutrinos at IceCube, complementary to the low-energy probes via neutrino oscillation and scattering experiments. 

Turning to the details of the scalar sector of the Zee model, in the Higgs basis~\cite{Davidson:2005cw} only the neutral component of $H_1$ acquires a vacuum expectation value (VEV) $\langle H_1^0\rangle =v\simeq 246.2$ GeV, while $H_2$ is parametrized as $H_2=(H_2^+, (H_2^0+iA^0)/\sqrt 2)$. The charged scalars $\{H_2^+,\eta^+\}$ mix  to give rise to the physical charged scalar mass eigenstates 
\begin{eqnarray}
        h^+ & \ = \ & \cos\varphi\, \eta^+ + \sin\varphi\, H_2^+ \, , 
        \quad 
        H^+ \ = \  -\sin\varphi \,\eta^+ + \cos\varphi \,H_2^+ \, ,
        \label{eq:charged}
    \end{eqnarray}
with the mixing angle $\varphi$  given by 
    \begin{equation}
        \sin{2\varphi} \ = \ \frac{-\sqrt{2} \ v \mu}{m_{H^+}^2-m_{h^+}^2}~,
        \label{mixphi}
    \end{equation}
    where $\mu$ denotes the coefficient of the cubic term $\mu H_1^iH_2^j\epsilon_{ij}\eta^-$ in the scalar potential. Here $i,j$ are $SU(2)_L$ indices. The Yukawa couplings in the lepton sector are given by the Lagrangian
    \begin{align}
        -{\cal L}_Y \ \supset \ &  f_{\alpha\beta}L_\alpha^i L_\beta^j \epsilon_{ij}\eta^+ + \widetilde{Y}_{\alpha \beta} \widetilde{H}_1^i L^{j}_\alpha   \ell_{\beta}^c \epsilon_{ij} 
         + Y_{\alpha\beta} \widetilde{H}_2^i L^{j}_\alpha  \ell_{\beta}^c \epsilon_{ij}  + {\rm H.c.} \, ,
        \label{eq:LYuk}
    \end{align}
    where $\{\alpha,\beta\}$ are flavor indices,  and $\widetilde{H}_a = i \tau_2 H^ \star_a$ ($a=1,2)$ with $\tau_2$ being the second Pauli matrix.

\begin{figure}[!t]
         \centering
         \includegraphics[scale=0.5]{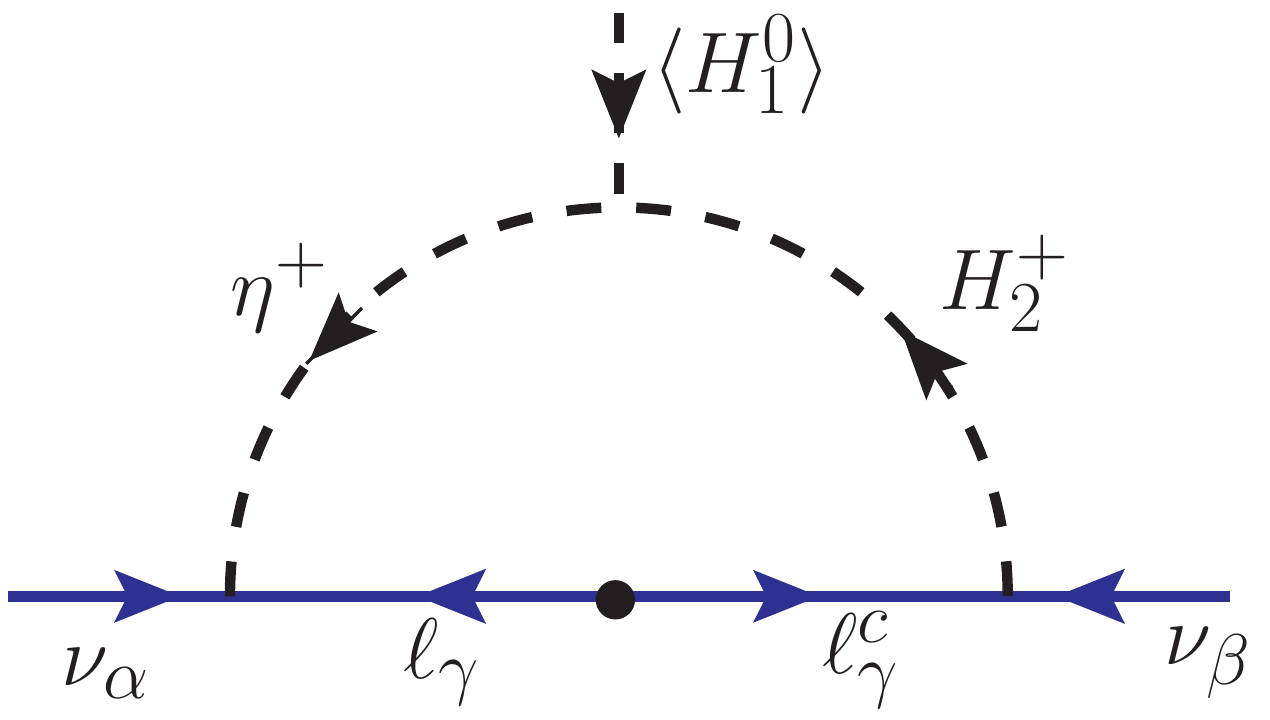}
         \caption{Neutrino mass generation at one-loop level in the Zee model~\cite{Zee:1980ai}. The dot ($\bullet$) on the fermion line indicates a lepton mass insertion utilizing the SM Higgs VEV.} 
         \label{fig:loopzee}
    \end{figure}

    The neutrino mass arises through a one-loop diagram shown in  Fig.~\ref{fig:loopzee}, and is given by 
    \begin{equation}
         M_\nu \ = \ \kappa \, (f M_\ell Y + Y^T M_\ell f^T) \, ,
         \label{nuMass}
    \end{equation}
where  $M_\ell=\widetilde Y v/\sqrt 2$ is the diagonal charged lepton mass matrix and $\kappa$ is a loop factor given by
    \begin{equation}
          \kappa \ = \ \frac{1}{16 \pi^2} \sin{2 \varphi} \log\left(\frac{m_{h^+}^2}{m_{H^+}^2}\right) \, .
          \label{kfactor}
    \end{equation}
It follows from  Eq.~\eqref{nuMass} that the product of the Yukawa couplings $f$ and $Y$ has to be relatively small in order to fit the neutrino oscillation data. This allows for one of these coupling matrices to have entries of order unity, while the other has much smaller entries. We shall choose  $Y\sim {\cal O}(1)$ and $f\ll 1$, which maximizes the neutrino NSI as well as resonance signals at IceCube in the model~\cite{Babu:2019mfe}. This is because the coupling $f_{\alpha\beta}$ is strongly constrained by processes such as muon decay, in the case where the scalars have sub-TeV masses~\cite{Herrero-Garcia:2017xdu}.

Note that since the model has two Higgs doublets, in general both doublets will couple to up-type and down-type quarks. If some of the leptonic Yukawa couplings $Y_{\alpha e}$ are of order unity, so that significant neutrino NSI can be generated, then the quark Yukawa couplings of the second Higgs doublet $H_2$ will have to be small. Otherwise if it couples to up and down quarks, chirality enhanced meson decays such as $\pi^+\to e^+\nu$  will occur with unacceptably large rates. It will also give unacceptably large contributions to beta decay $n\to p e^- \bar{\nu}$. Similarly, for couplings to second-generation quarks, there will be constraints from $D$-meson decays like $D^+\to e^+\nu$. And for third-generation quark couplings, there will be large modifications to the top and bottom decay widths. Therefore, we assume that the second Higgs doublet $H_2$ has negligible couplings to the quarks. Thus we take $H_2$ to be leptophilic in our analysis.

Our main focus for IceCube phenomenology is the resonant production of one or both of the charged scalars present in the model.  This can occur with measurable strength provided that these scalars have masses of order 100 GeV and the Yukawa couplings $Y_{\alpha e}$ are of order one for $\alpha = e,\mu$ or $\tau$. The light charged scalar scenario is confronted with several theoretical and experimental constraints, such as avoidance of charge breaking minima, consistency with electroweak precision tests, charged lepton flavor violation (cLFV), collider constraints from LEP and LHC, lepton universality tests and monophoton limits. It has been shown~\cite{Babu:2019mfe, Cao:2017ffm} that both $h^+$ and $H^+$ charged scalars can be as light as $\sim 100$ GeV, while satisfying all these constraints. The most stringent constraints come from direct searches at LEP experiment, which are applicable as long as $Y_{\alpha e}\neq 0$ for any flavor $\alpha$. 
There are more stringent constraints from lepton universality tests at LEP in $W$ decays~\cite{LEP:2003aa} if $Y_{ee}\neq 0$, which would restrict the charged scalar masses to above 130 GeV~\cite{Babu:2019mfe}. We shall take $Y_{\tau e}\neq 0$ and $Y_{\alpha \tau}\neq 0$ for $\alpha=e$ or $\mu$, which can satisfy all constraints for $m_{h^+}=100$ GeV, and at the same time, allows for the largest NSI effect as well as resonance signals at IceCube. 

\subsection{Event Spectrum at IceCube} \label{sec:eventzee}
 Eq.~\eqref{eq:LYuk} contains the interaction terms 
\begin{align}
    {\cal L}_Y \ \supset \ Y_{\alpha\beta}(h^-\sin\varphi + H^-\cos\varphi)\nu_\alpha \ell_\beta^c + {\rm H.c.}
    \label{eq:LYuk1}
\end{align}
For $\beta=e$, this will induce anti-neutrinos scattering off electrons,  $\overline{\nu}_\alpha + e^- \rightarrow h^- \rightarrow \overline{\nu}_\beta + \ell^-$ as well as
$\overline{\nu}_\alpha + e^- \rightarrow H^- \rightarrow \overline{\nu}_\beta + \ell^-$ where $\ell^-$ in the final state may be $e^-, \mu^-$ or $\tau^-$.  For anti-neutrino energy $E_\nu=m_{h^-(H^-)}^2/2m_e$, the cross sections for these processes is resonantly enhanced  (Zee-burst) at IceCube. 
The Zee-burst does not interfere with the Glashow resonance (for $\alpha = e$) since the helicity of the electrons is opposite in the two cases.   Thus, depending on how close the masses of the $h^-$ and $H^-$ fields are, we would expect additional resonance peaks in the IceCube energy spectrum. We will consider two benchmark scenarios: (i) $m_{h^-}\approx m_{H^-}$, so that the two peaks are indistinguishable, i.e. they fall in the same energy bin, and (ii) $\Delta m_h\equiv m_{H^-} -m_{h^-}=30$ GeV, so that the two resonance peaks are distinguishable, i.e. they fall in different energy bins. It is worth noting that we cannot take $\Delta m_h$ exactly zero, since this would result in the neutrino mass being zero [cf.~Eq.~\eqref{kfactor}]. 

To get the event spectrum, we compute the number of events in a given energy bin $i$ as
\begin{align}\label{eq:reconst}
N_i \  = \ T \int_{0}^{4\pi} d\Omega \int_{E_i^{\rm min}}^{E_i^{\rm max}} dE \sum_{\alpha} \Phi_{\nu_\alpha}(E)  A_{\nu_\alpha}(E,\Omega) \, .
\end{align}
Here $T$ is the exposure time which we take to be $T_0=2653$ days, corresponding to 7.5 years of live data taking at IceCube~\cite{Aartsen:2019kpk}; $\Omega$ is the solid angle of coverage which we integrate over the whole sky (taking into account the attenuation for incoming neutrinos below the horizon); $E$ is the electromagnetic-equivalent deposited energy which is an approximately linear function of the incoming neutrino energy $E_\nu$~\cite{Palladino:2018evm}; the limits of the energy integration $E_i^{\rm min}$ and $E_i^{\rm max}$ determine the $i$th deposited energy bin over which the expected number of events is being calculated;  $\Phi_{\nu_\alpha}(E)$ is the differential astrophysical neutrino+anti-neutrino flux for a given flavor $\alpha$ (which is summed over), for which we use the power-law flux given by Eq.~\eqref{eq:flux}; and $A_{\nu_\alpha}$ is the effective area per energy per solid angle for flavor $\alpha$, which includes the effective neutrino-matter cross section, number density of target electrons/nucleons and acceptance rates for the shower/track events. The new interactions present in Eq.~\eqref{eq:LYuk1} would modify the (anti)neutrino-electron cross section, which would lead to a modification of the effective area. For neutrino interactions within the SM, we use the flavor-dependent effective area integrated over solid angle from Ref.~\cite{Aartsen:2017mau} (for 2078 days of IceCube data), and increase the acceptance by $67\%$ to correspond to 2653 days of data~\cite{Aartsen:2019epb}. In order to include the non-standard interaction of Eq.~\eqref{eq:LYuk1}, we rescale the effective area by taking the ratio of the two cross sections (with and without the new interactions), assuming that the acceptance remains the same. 

\begin{figure}[t!]
\centering
\includegraphics[width=0.9\textwidth]{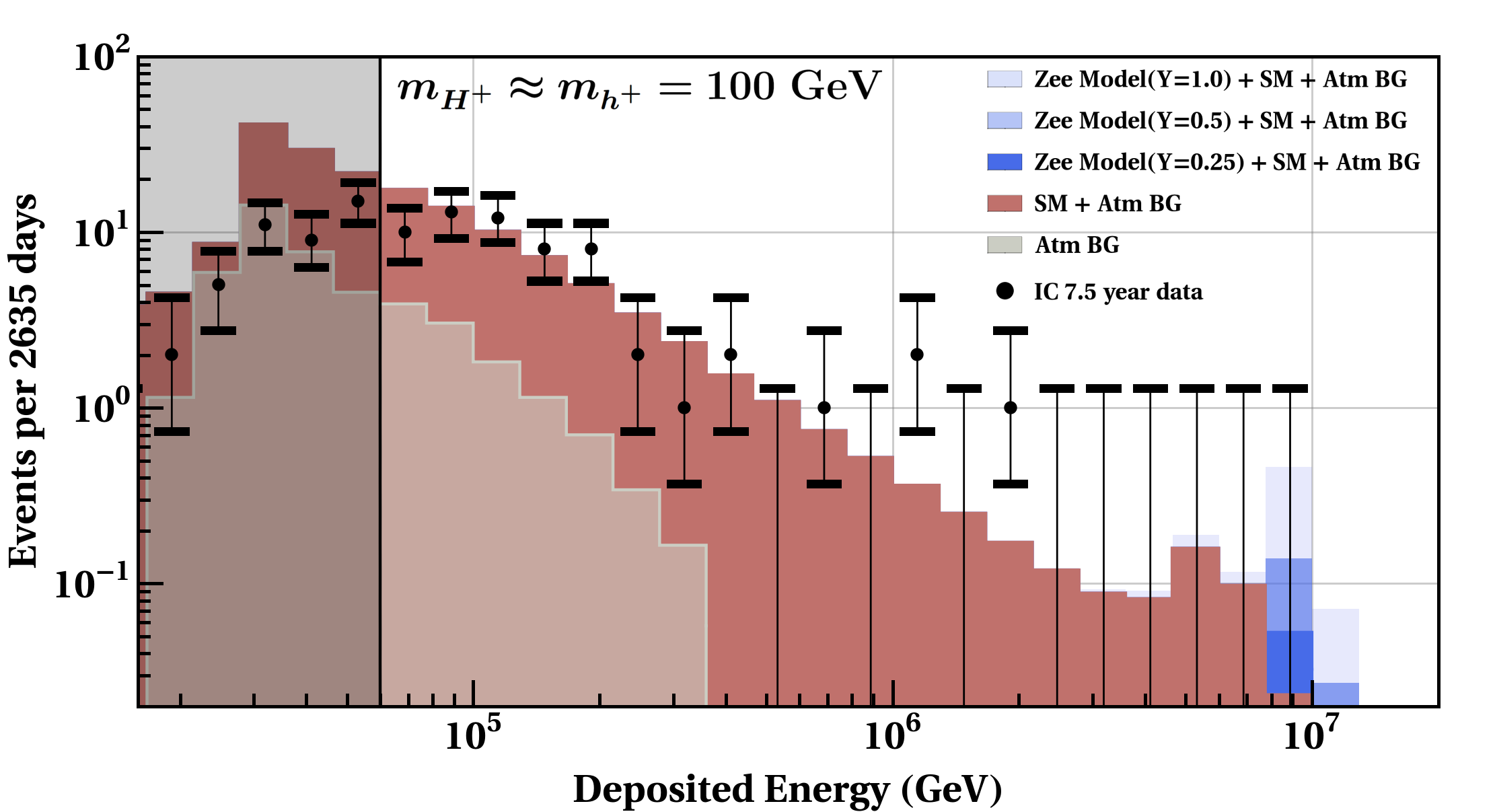}
\caption{Reconstructed UHE neutrino event spectra at IceCube for the expected atmospheric background (gray), SM best-fit with a single-component astrophysical flux (red) and the Zee model with $m_{h^+}\approx m_{H^+}=100$ GeV, $\varphi=\pi/4$ and $Y_{\tau e}=1, 0.5, 0.25$ (light, medium and dark blue, respectively), all compared with the 7.5-year IceCube data (black dots with error bars). The data points below 60 TeV (vertical black-shaded region) are not included in the IceCube HESE analysis we are using here. Figure reproduced from Ref.~\cite{Babu:2019vff}. \label{fig:spectra}
}
\end{figure}

In the absence of new interactions, neutrinos interact with nucleons via charged- and neutral-current processes of the SM. In the energy range of interest for UHE neutrinos at IceCube, the corresponding DIS cross sections is given approximately by~\cite{Gandhi:1995tf} 
\begin{align}
\sigma_{\nu(\bar\nu)N}^{\rm CC} \ \approx \  3\sigma_{\nu(\bar\nu) N}^{\rm NC} \ \simeq \  2.7\times 10^{-36} {\rm cm^2}\left(\frac{E_{\nu}}{{\rm GeV}}\right)^{0.4}.
\end{align}
The (anti)neutrino-electron interactions are subdominant for all energies, except in the range of 4.6--7.6 PeV. In this range $\bar{\nu}_e$--$e^-$ interaction cross section is enhanced owing to the Glashow resonance~\cite{Glashow:1960zz}.  In the neighborhood of  this resonance, the cross section can be expressed by a Breit–Wigner formula~\cite{Barger:2014iua}:
\begin{align}\label{eq:GlashowXsection}
    \sigma_{\rm Glashow}(s) 
\ = 
\  & 24\pi\,\Gamma_W^2\, {\rm BR}(W^-\to\bar\nu_e e^-){\rm BR}(W^-\to{\rm had}) 
      \frac{s/m_W^2}{(s-m_W^2)^2+(m_W\Gamma_W)^2} \, ,
\end{align}
where $s=2m_{e}E_{\nu}$ and $\Gamma_{W}=2.1$ GeV is the width of the $W$ boson with ${\rm BR}(W^-\to\bar\nu_e e^-)=10.7\%$ and ${\rm BR}(W^-\to{\rm had})=67.4\%$~\cite{ParticleDataGroup:2020ssz}. 
At resonance, the cross section arising from Eq.~\eqref{eq:GlashowXsection} is $\sigma_{\rm Glashow}(E_{\nu}=6.3 \: {\rm PeV})=3.4\times10^{-31}\rm\: cm^2$, which is roughly 200 times larger than $\sigma_{\nu(\bar\nu)N}^{\rm CC}(E_{\nu}=6.3\: {\rm PeV})\approx 1.4\times10^{-33}\:\rm cm^2$.
However, owing to the narrowness of the resonance and the $E_\nu^{-\gamma}$ power-law dependence on the neutrino energy, the ratio of the reconstructed events between the resonance-induced $\bar{\nu}_e$-$e$ and non-resonant $\nu(\bar \nu)$-$N$ interactions is not as pronounced in the event spectrum. We display this by the red-shaded histograms in Fig.~\ref{fig:spectra}. As an example,
for $E_{\nu}>4$ PeV, $N_{\rm Res}/N_{\rm non-Res}\sim 2$, which results in a total of about 0.3 events in the Glashow bin for the IceCube best-fit flux. This is consistent within $1\sigma$ with the IceCube observation of one Glashow event~\cite{IceCube:2021rpz}. Also shown in Fig.~\ref{fig:spectra} by the gray-shaded region is the total expected atmospheric background (from atmospheric muons and neutrinos, as well as the charmed meson contribution). The 7.5 year IceCube data~\cite{IceCube:2020wum} is shown by the black dots with $1\sigma$ error bars.
The low-energy cutoff for the HESE analysis is indicated by the vertical line at 60 TeV. In other words, the bins below this energy are not taken into account in the fitting process for obtaining the best-fit astrophysical neutrino flux.

In presence of new light charged scalars, as in the Zee model of radiative neutrino mass, we expect a new resonance in the scattering process $\bar\nu_\alpha e^- \to X^- \to \rm anything$ (where $X^-=h^-,H^-$ for the Zee model) with a Breit-Wigner form for the cross section similar to Eq~\eqref{eq:GlashowXsection}:
\begin{align}
   \sigma_{\rm Zee} (s) \ = \ & 8\pi\,\Gamma_X^2 \,{\rm BR}(X^-\to\bar\nu_\alpha e^-) {\rm BR}(X^-\to{\rm all}) 
    \frac{s/m_X^2}{(s-m_X^2)^2+(m_X\Gamma_X)^2} \, ,
    \label{eq:ZeeXsection}
\end{align}
where $\Gamma_X=\sum_{\alpha\beta}|Y_{\alpha\beta}|^2{\rm sin^2 \varphi}\:m_X/16\pi$ is the total decay width of $X$. There is an overall relative suppression by a factor of 1/3 in Eq. (\ref{eq:ZeeXsection}), compared to Eq.~\eqref{eq:GlashowXsection}, which arises due to the difference in the degrees of polarization between scalar and vector bosons. 

We consider a benchmark case with $m_{h^-}\approx m_{H^-}=100$ GeV in Fig.~\ref{fig:spectra}. With this choice the  two new resonances due to $h^-$ and $H^-$ coincide, and thus, maximize the effect in the bin containing the resonance energy $E_\nu=m_{h^-}^2/2m_e$, as shown by the light, medium and dark blue-shaded histograms respectively corresponding to three illustrative values of $Y_{\tau e}=1,0.5,0.25$. The events due to the Zee scalar resonance mostly populate the energy bins between 7.6 and 12.9 PeV, which contrasts with those characterized by the Glashow resonance bin (4.6--7.6 PeV). As expected from Eq.~\eqref{eq:ZeeXsection}, the new resonance effect is more pronounced for larger Yukawa couplings. Here for illustration we have chosen the maximal mixing $\varphi=\pi/4$ and ${\rm BR}(h^-\to \bar{\nu}_\tau e)=60\%$, ${\rm BR}(h^-\to \bar{\nu}_\beta \tau)=40\%$ (with $\beta=e$ or $\mu$) for a fixed $Y_{\tau e}$ noted above. The Yukawa coupling $Y_{\beta\tau}$ is chosen to reproduce these branching ratios, while all other Yukawa couplings $Y_{\alpha\beta}$ are considered to be negligible so that the cLFV constraints are satisfied~\cite{Babu:2019mfe}. Not that as the mass difference  $\Delta m_h\equiv m_{H^-}- m_{h^-}$ increases, the two peaks begin to populate distinct bins, although the impact is more evident in the lowest resonance energy bin due to the dropping power-law flux.

\begin{figure*}[t!]
  \centering
 \includegraphics[width=0.49\textwidth]{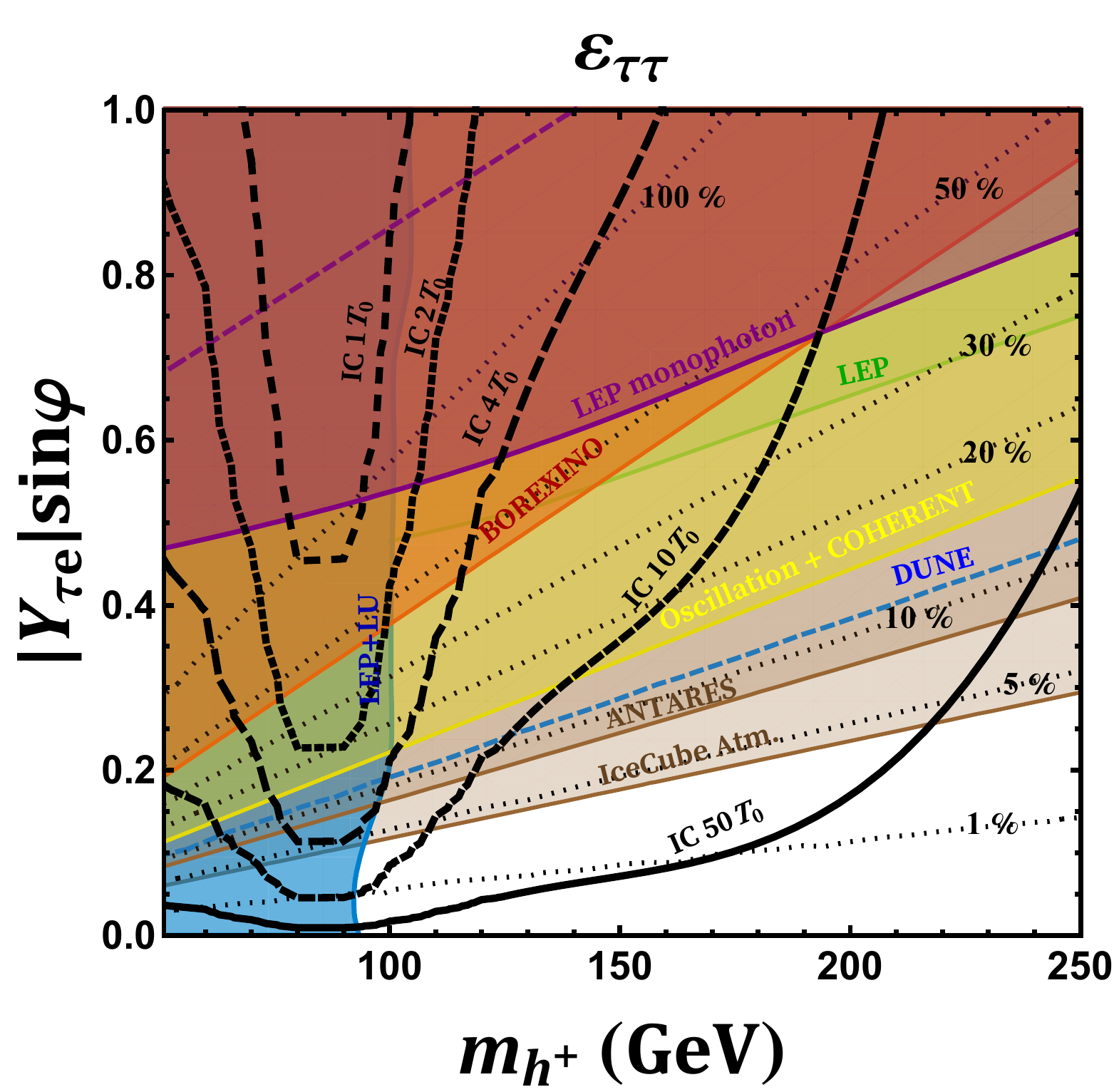}
\includegraphics[width=0.49\textwidth]{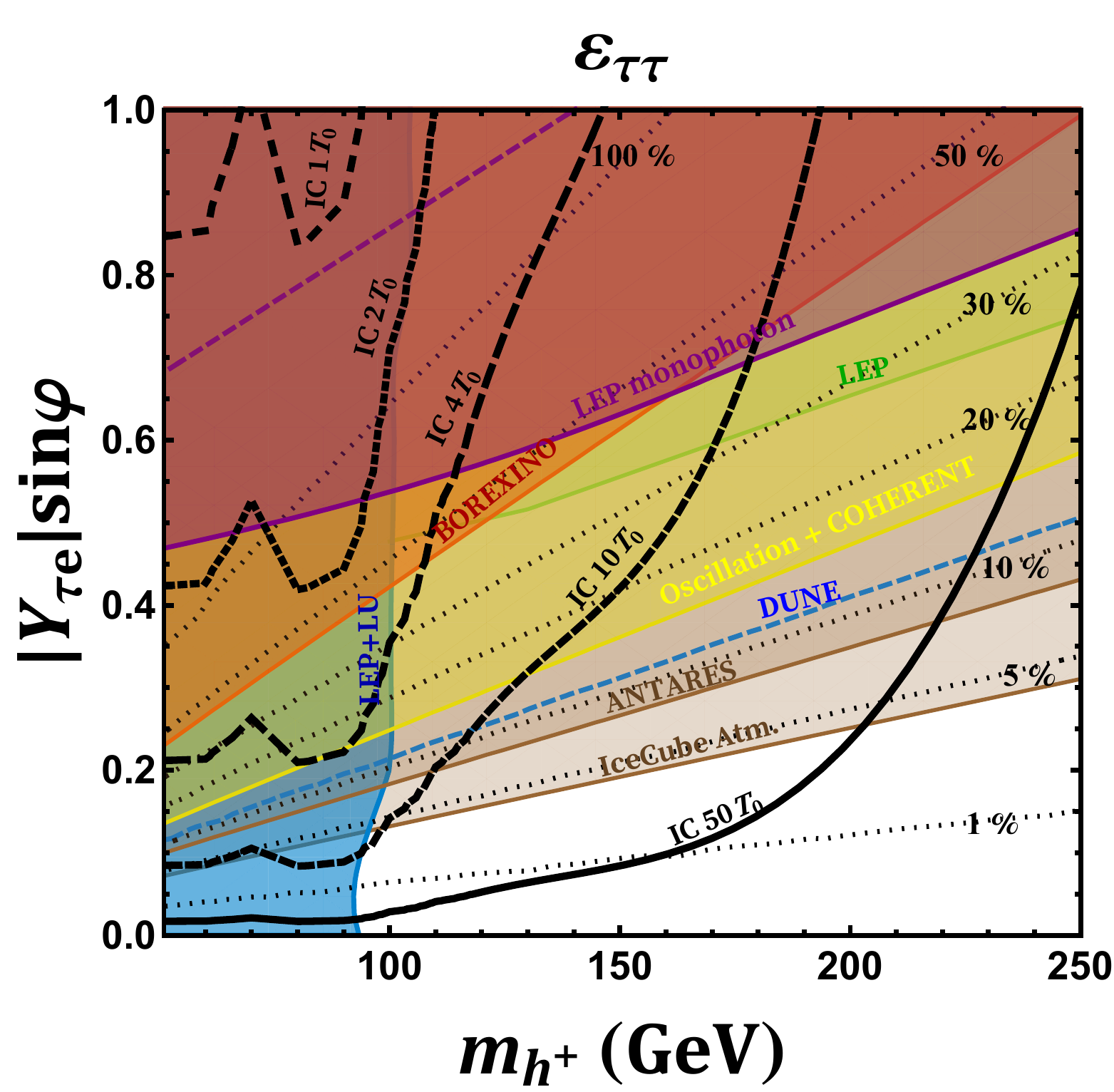}
  \caption{IceCube sensitivity contours (corresponding to one expected event in the resonance energy bins combined)  for the parameter space relevant for $\varepsilon_{\tau\tau}$ are shown by thick black curves, for different exposure times (in terms of the current exposure $T_0=2653$ days). The left panel represents $m_{h^+}\approx m_{H^+}$, whereas the right panel represents $m_{H^+}-m_{h^+}=30$ GeV. 
  The thin dotted contours indicate the predictions for $\varepsilon_{\tau\tau}$. 
  The shaded regions are excluded: blue-shaded by direct LEP searches~\cite{lepsusy, Abbiendi:2013hk} and lepton universality (LU) tests in tau decays~\cite{ParticleDataGroup:2020ssz}; green-shaded by LEP dilepton searches~\cite{LEP:2003aa, Abbiendi:2003dh}; purple-shaded  (dashed) by LEP monophoton searches off (on) $Z$-pole~\cite{Acciarri:1998vf, Achard:2003tx}; red-shaded by BOREXINO~\cite{Agarwalla:2019smc},  yellow-shaded by global fit to neutrino oscillation plus COHERENT data~\cite{Coloma:2019mbs}, and light (dark) brown by IceCube~\cite{IceCubeCollaboration:2021euf} (ANTARES~\cite{Albert:2021sfx}) atmospheric neutrino data. For more details on these exclusion regions, see Ref.~\cite{Babu:2019mfe}. For comparison, we also show the future DUNE sensitivity in blue dashed line~\cite{Chatterjee:2021wac}. Figure updated from Ref.~\cite{Babu:2019vff}.
  }
  \label{fig:NSI}
\end{figure*}

It is clear from Fig.~\ref{fig:spectra} that for a given charged scalar mass $m_{h^-}$, the Yukawa coupling $Y_{\tau e}$ cannot be made arbitrarily large without spoiling the best-fit to the observed IceCube HESE data. This enables us to derive new IceCube constraints on charged scalar masses versus couplings which we show in Fig.~\ref{fig:NSI} by the thick black contours. The curve labeled `IC 1$T_0$' corresponds to the parameter set that would result in a single event  when summed over the last three bins considered by IceCube best-fit ($4.6<E_\nu/{\rm PeV}<10$) with the current exposure  of $T_0=2653$ days~\cite{IceCube:2020wum}. The other contours correspond to increased exposures of $2T_0$, $4T_0$, $10T_0$ and $50 T_0$ respectively, for the same set ot parameters in Eq.~\eqref{eq:reconst} and assuming that the acceptance remains the same. 
The left panel corresponds to $m_{h+} \approx m_{H^+}$ and the right panel is for $m_{H^+}-m_{h^+}=30$ GeV. This explains the appearance of one `dip' in the left panel (corresponding to one resonance for $h^-$ and $H^-$ combined) and two `dips' in the right panel (corresponding to two distinct resonances for $h^-$ and $H^-$).  

\subsection{Probing NSI} 
The Yukawa interactions of Eq.~\eqref{eq:LYuk1} would lead to neutrino NSI with electrons, given by~\cite{Babu:2019mfe}
\begin{equation}
    \varepsilon_{\alpha \beta}  \ = \ \frac{Y_{\alpha e } Y_{ \beta e}^{\star}}{4 \sqrt{2}G_F} \left(\frac{ \sin^2{\varphi}}{ m_{h^+}^2}  +   \frac{\cos^2{\varphi} }{m_{H^+}^2}\right) \, ,
         \label{nsieqntot}
\end{equation}
where $G_F$ is the Fermi coupling constant. We show in Fig.~\ref{fig:NSI} the predictions for $\varepsilon_{\tau\tau}$ by thin black dotted contours.  
Here again we have assumed maximal mixing among the charges scalars with $\varphi=\pi/4$ to illustrate the largest possible NSI. The excluded shaded regions arise from various constraints: red-shaded by BOREXINO~\cite{Agarwalla:2019smc};  orange-shaded by global fit to neutrino oscillation plus COHERENT data~\cite{Coloma:2019mbs}; brown-shaded by IceCube atmospheric neutrino data~\cite{IceCubeCollaboration:2021euf}; blue-shaded by direct LEP searches~\cite{lepsusy, Abbiendi:2013hk} and lepton universality (LU) tests in tau decays~\cite{ParticleDataGroup:2020ssz}; green-shaded by LEP dilepton searches~\cite{LEP:2003aa, Abbiendi:2003dh}; and purple-shaded with solid (dashed) boundary by LEP monophoton searches off (on) $Z$-pole~\cite{Acciarri:1998vf, Achard:2003tx}. 
For a detailed discussion of these exclusion regions, see Ref.~\cite{Babu:2019mfe}. Note that the IceCube atmospheric neutrino data only constrains $|\varepsilon_{\tau \tau}-\varepsilon_{\mu\mu}|< 4.2\%$~\cite{IceCubeCollaboration:2021euf}, which in the Zee model is equivalent to a bound on $\varepsilon_{\tau\tau}$ itself, because both $\varepsilon_{\tau\tau}$ and  $\varepsilon_{\mu\mu}$ cannot be large simultaneously due to stringent cLFV constraints. A slightly weaker bound on $\varepsilon_{\tau \tau}$ itself has been recently reported by the ANTARES collaboration~\cite{Albert:2021sfx}: $|\varepsilon_{\tau \tau}|<8.1\%$ at 90\% CL (assuming normal ordering). A similar analysis with other $\varepsilon_{\alpha\beta}$ would however lead to weaker limits, less than a few \%~\cite{Babu:2019mfe}, and hence, are not so promising for IceCube HESE sensitivity.

It follows from Fig.~\ref{fig:NSI} that the existing limits on NSI are more stringent than the current sensitivity of high-energy IceCube data. Nevertheless, the (non)observation of a resonance-like signature in the future IceCube HESE data might serve as  a complementary probe of the allowed NSI parameter space within the Zee model of neutrino mass, which can even supersede the future DUNE sensitivity (shown by the blue dashed line~\cite{Chatterjee:2021wac}) and $e^+e^-$ collider sensitivity~\cite{Liao:2021rjw}. We should point out that an exposure of $10T_0$ (with $T_0$ being roughly 7.5 years) does not necessarily require 75 years of IceCube running,  since a variety of  unforeseen factors could improve the conservative projected IceCube limits presented here in a non-linear fashion. For example, the future data in all the bins may not scale proportionately to the current data and may turn out to be in better agreement with the SM prediction, which would restrict even further any room for new physics contributions. Similarly, there is a good chance that the energy-dependent acceptance rate would improve in the future (as it did by 67\% from two to seven years of data~\cite{Aartsen:2019epb}), thereby increasing the effective area, and hence, the `effective' exposure time at a rate faster than linear. Finally, the proposed IceCube-Gen2 upgrade with 10 km$^3$ detector  volume~\cite{IceCube-Gen2:2020qha} 
might increase the total effective exposure by about an order of magnitude. At the very least, combining IceCube with ANTARES~\cite{ANTARES:2011hfw} or future KM3NeT~\cite{Adrian-Martinez:2016fdl} or P-ONE~\cite{P-ONE:2020ljt} experiments could increase the effective exposure by at least a factor of few.  

Before concluding this section, we remark that for heavier charged scalars, the resonance energy will be shifted to higher values for which IceCube will become less sensitive, assuming as we do an isotropic power-law spectrum of the neutrino flux.  However, if there are powerful transient sources of UHE neutrinos, then IceCube, as well as current and next-generation radio-Cherenkov neutrino detectors, such as ARA~\cite{Allison:2011wk}, ARIANNA~\cite{Barwick:2014pca}, ANITA~\cite{Gorham:2019guw}, 
PUEO~\cite{PUEO:2020bnn}, GNO~\cite{Avva:2016ggs}, RNO~\cite{Aguilar:2019jay} and GRAND~\cite{GRAND:2018iaj}, could be sensitive to charged scalars without neutrino-electron couplings up to around a TeV (corresponding to the resonance energy of EeV), as might occur for e.g. in left-right symmetric model~\cite{Boyarkin:2017rte}. For a recent study on the tau neutrino telescope, see Ref.~\cite{Huang:2021mki}. The prospect of a larger flux at higher energies, along with improved energy resolution of the IceCube detectors, might aid in discriminating between the degenerate vs non-degenerate charged-scalar mass spectrum of the Zee model by exploiting the `dip' features in Fig.~\ref{fig:NSI}.

\section{Scalar Leptoquarks at IceCube}\label{sec:LQ}

\begin{figure}[t!]
    \centering
    \includegraphics[scale=0.5]{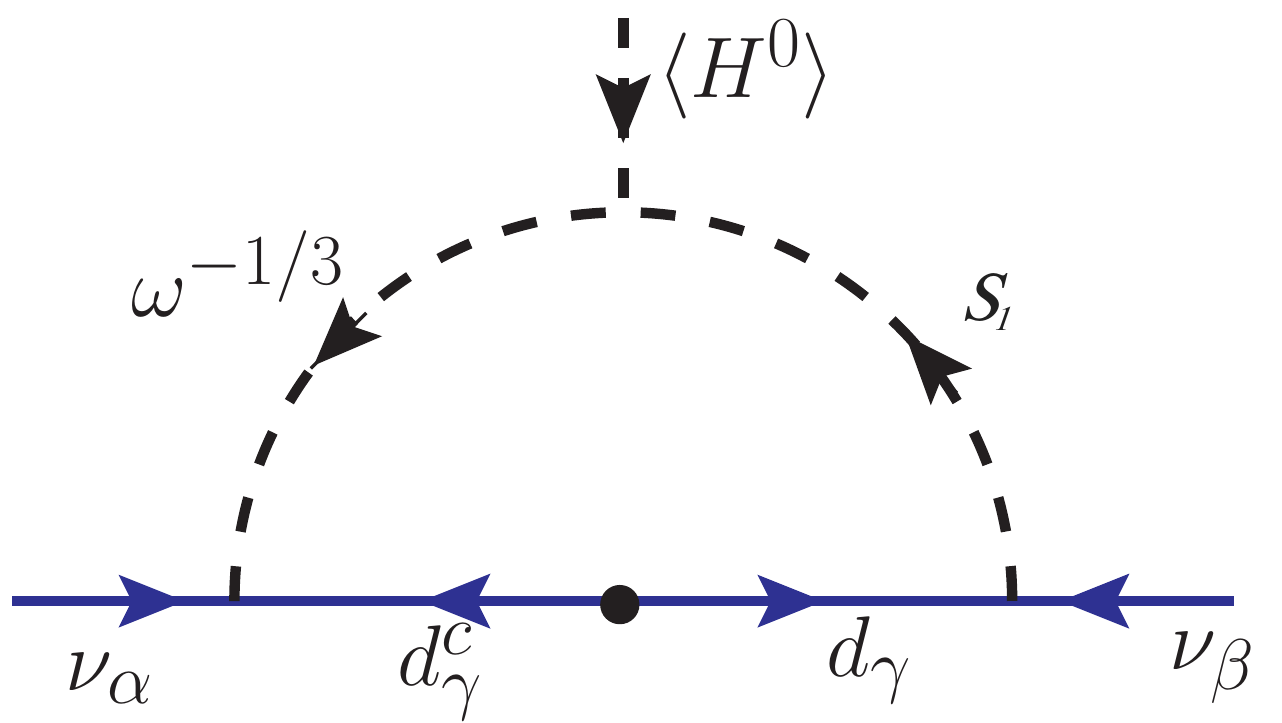}
    \caption{One-loop diagram inducing neutrino mass in a colored variant of the Zee model with $SU(2)_L$-singlet ($S_1$) and doublet ($\widetilde{R}_2$) leptoquarks. The dot ($\bullet$) on the fermion line indicates a quark mass insertion utilizing  the SM Higgs VEV.}
    \label{fig:coloredzee}
\end{figure}

As our next example, we take the colored variants of the Zee model~\cite{Hall:1983id} where the charged scalars in Fig.~\ref{fig:loopzee} are replaced by  scalar leptoquarks~\cite{Babu:2019mfe}; see Fig.~\ref{fig:coloredzee} for an example. We adopt the notation introduced in Ref.~\cite{Buchmuller:1986zs}. As shown in Table~\ref{tab:glossory}, a resonance feature at IceCube can be induced by all four types of scalar leptoquarks, namely, $S_1$, $S_3$, $R_2$ and $\widetilde{R}_2$.   Considering one leptoquark at a time, we examine all the interaction terms in the given leptoquark Lagrangian that could give such a resonance. The relevant Lagrangian terms are given by 
\begin{align}
    {\cal L}_Y^{S_1} & \supset \lambda_{\alpha\beta}(\nu_\alpha d_\beta-\ell_\alpha u_\beta)S_1 +{\rm H.c.} \, , \label{eq:LS1} \\
    {\cal L}_Y^{S_3} & \supset \lambda'_{\alpha\beta}L_\alpha Q_\beta S_3 =  \lambda'_{\alpha\beta}\left[\ell_\alpha d_\beta \rho^{4/3}-\frac{1}{\sqrt 2}(\nu_\alpha d_\beta+\ell_\alpha u_\beta)\rho^{1/3}+\nu_\alpha u_\beta \rho^{-2/3}\right] +{\rm H.c.} \, , \\
    {\cal L}_Y^{R_2} & \supset \lambda''_{\alpha\beta}L_\alpha^i u_\beta^c R_2^j \epsilon_{ij} =  \lambda''_{\alpha\beta}(\nu_\alpha u^c_\beta\delta^{2/3}-\ell_\alpha u_\beta^c \delta^{5/3}) +{\rm H.c.}  \, , \\
    {\cal L}_Y^{\widetilde{R}_2} & \supset \lambda'''_{\alpha\beta}L_\alpha^i d_\beta^c \widetilde{R}_2^j \epsilon_{ij} =  \lambda'''_{\alpha\beta}(\nu_\alpha d^c_\beta\omega^{-1/3}-\ell_\alpha d_\beta^c \omega^{2/3}) +{\rm H.c.} \label{eq:LR2t}
\end{align}
The same leptoquark couplings $\lambda_{\alpha\beta}$ that govern the neutrino-quark interactions (and hence the IceCube signal) also induce charged lepton-quark interactions. Both of these final states have been searched for the LHC, which impose stringent constraint on the leptoquark couplings, which we will discuss in Sec.~\ref{sec:LQLHC}. 

Neutrino masses are induced via one-loop diagrams involving two leptoquarks. For instance, for the example shown in Fig.~\ref{fig:coloredzee}, the mass matrix is given by
\begin{equation}
    M_\nu \ = \ \frac{3 \sin 2\alpha}{32 \pi^2} \log \left(\frac{m_1^2}{m_2^2}\right) (\lambda''' M_d \lambda^T + \lambda''' M_d \lambda^T)\, ,
    \label{eq:Mnucz}
\end{equation}
where $M_d$ is the diagonal down-type quark mass matrix, $m_{1,2}$ are the eigenvalues of the mixed singlet-doublet leptoquark mass matrix:
\begin{equation}
    m_{1,2}^2 \ = \ \frac{1}{2} \left[ m_\omega^2 + m_{S_1}^2 \mp \sqrt{(m_\omega^2 - m_{S_1}^2)^2 + 4 \mu^2 v^2} \right] \, ,
    \label{eq:mX12}
\end{equation}
$\mu$ denotes the coefficient of the cubic scalar coupling $\mu H^\dag \widetilde{R}_2 S_1$ and 
$\alpha$ is the mixing angle between the singlet and $\omega^{-1/3}$-component of the doublet:
\begin{equation}
    \tan 2\alpha \ = \ \frac{-\sqrt{2} \, \mu v}{m_{S_1}^2 - m_\omega^2}\,.
    \label{eq:tan2a}
\end{equation}
Acceptable neutrino masses and mixing can be generated for a variety of parameters.  Note that the induced $M_\nu$ is proportional to the down-quark masses, the largest being $m_b$. In the spirit of maximizing neutrino NSI, which are induced by either the $\omega^{-1/3}$ or the $S_1$ field, without relying on their mixing, we shall adopt a scenario where either $\lambda_{\alpha\beta}$ or $\lambda'''_{\alpha\beta}$ is of order one, while the other one is much smaller than one, which would realize small neutrino masses. Similar choices can be made for the other combinations of the leptoquarks; therefore, we will consider one sizable leptoquark coupling at a time for our IceCube analysis below, but it is generically applicable to all leptoquark-induced radiative neutrino mas models~\cite{Babu:2019mfe}. 

\subsection{Event Spectrum at IceCube} \label{sec:eventLQ}
  \begin{figure}[t!]
    \centering
    $$
     \includegraphics[width=.9\textwidth]{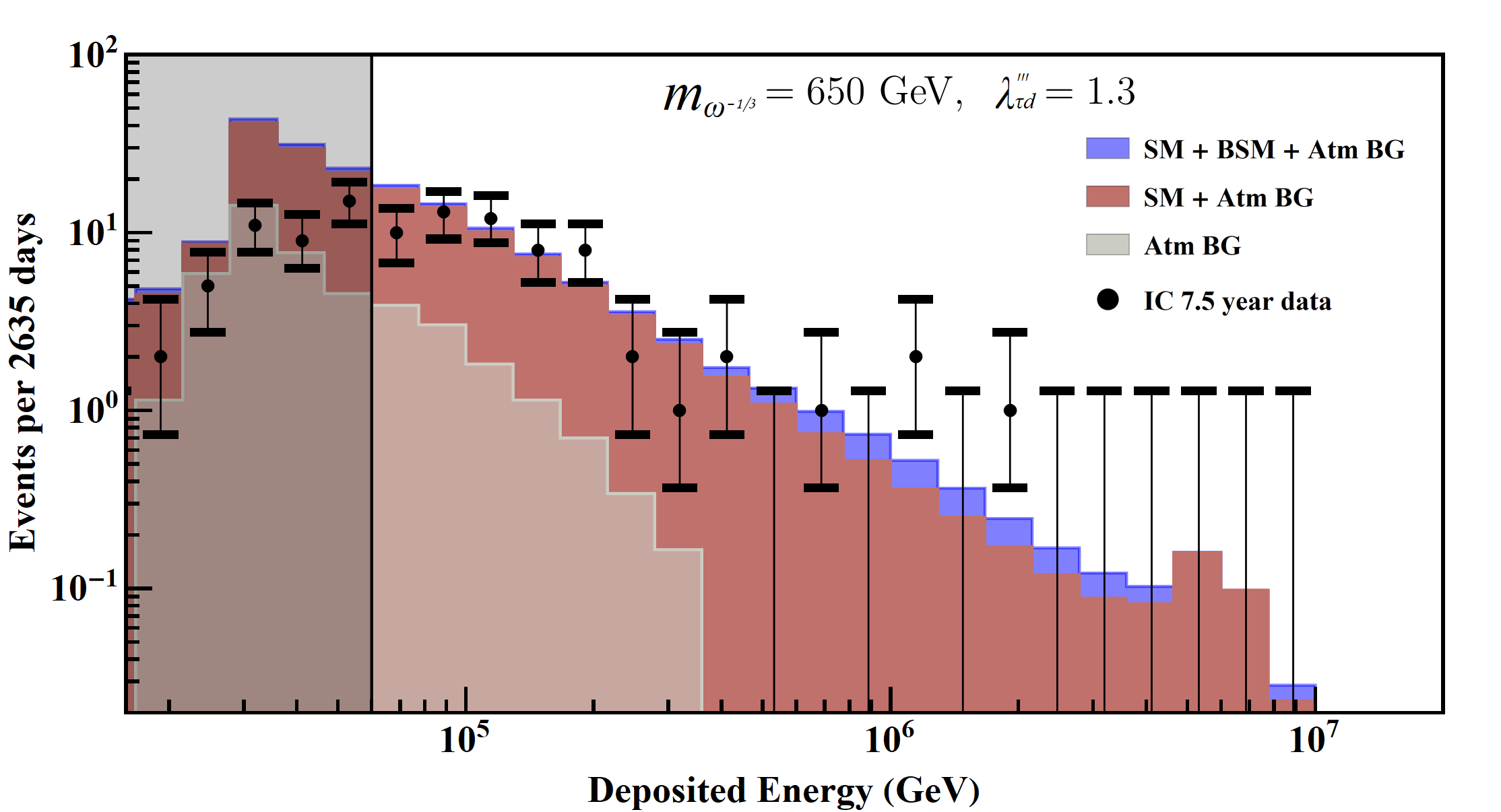}
     $$
    \caption{Reconstructed event spectra for the expected atmospheric background (gray), SM best-fit with a single-component astrophysical flux (red) and for the $\widetilde{R}_2$ leptoquark model (blue) with mass 650 GeV for the $\omega^{-1/3}$ component and Yukawa coupling $\lambda'''_{\tau d}=1.3$ as an example, all compared with the 7.5-year IceCube data. The data points below 60 TeV (as shown by the vertical black-shaded band) are not included in our IceCube HESE analysis. 
    }
    \label{fig:LQspectrum}
\end{figure}

    The leptoquark interactions with neutrinos and quarks in Eqs.~\eqref{eq:LS1}-\eqref{eq:LR2t} could also give distinct resonance features in the IceCube HESE data. The basic idea is that a (sub) TeV-scale leptoquark can be resonantly produced by the interaction of multi-TeV astrophysical neutrinos with Earth matter (protons/neutrons), as shown in the bottom row of Fig.~\ref{fig:feyn_spectra}. The resonance condition is satisfied for the incoming neutrino energy $E_\nu=m_{\rm LQ}^2/2m_Nx$, where $m_N$ is the nucleon mass and $x=Q^2/2m_NE_\nu'$ is the Bjorken scaling variable representing the initial parton momentum fraction, with $Q^2$ being the invariant momentum transfer and $E_\nu'$ is the energy loss in the laboratory frame. Due to the spread in $x\in [0,1]$, the resonance peak will be broadened and shifted above the threshold value $E_\nu^{\rm th}=m_{\rm LQ}^2/2m_N\sim 500~{\rm TeV}$ for $m_{\rm LQ}=1$ TeV, unlike the Zee scalar case where the resonance was much narrower (see Fig.~\ref{fig:spectra}). This can be clearly seen in Fig.~\ref{fig:LQspectrum} where the broad resonance contributes to excess events in multiple bins, way  above the threshold of $E_\nu^{\rm th}=211$ TeV for $m_{\rm LQ}=650$ GeV chosen in this example. We have also fixed the Yukawa coupling $\lambda_{\tau d}=1.3$ and all other couplings zero in this illustration, and used the {\sc NNPDF3.1}~\cite{NNPDF:2017mvq} PDF sets for the neutrino-nucleon cross-section calculation. The rest of the event spectrum generation procedure is the same as in Sec.~\ref{sec:eventzee}.  The relevant differential cross-sections can be found e.g. in Ref.~\cite{Collins:2018jpg} which were numerically integrated to obtain the total cross section. 
    
    From Fig.~\ref{fig:LQspectrum}, it is clear that  for a given leptoquark mass, the Yukawa coupling cannot be arbitrarily large without spoiling the best-fit to the observed IceCube HESE data. Similar to what was done in Sec.~\ref{sec:eventzee}, we can use this fact to derive new IceCube constraints in the leptoquark mass and Yukawa plane. This is shown by the black-shaded region to the top left corner of each panel in Fig.~\ref{fig:LQ}, corresponding to the 95\% CL exclusion limit derived from the existing IceCube HESE data ($1T_0$) in a bin-by-bin analysis. The future IceCube sensitivity is shown by the thick red contours. The curve labeled `IC 10$T_0$' represents the parameter set which would give rise to one excess event when summed over all the bins considered by IceCube best-fit ($4.6<E_\nu/{\rm PeV}<10$) with 10 times the current exposure $T_0=2653$ days~\cite{IceCube:2020wum}. Similarly, the $100 T_0$ curve corresponds to 100 times the current exposure, keeping the other parameters in Eq.~\eqref{eq:reconst} the same.  Unfortunately, even with 100 times the current exposure, the parameter space sensitive to a leptoquark resonance at neutrino telescopes is already excluded by the LHC constraints on leptoquarks, which is not very surprising, given the fact that LHC being a hadron collider usually provides the strongest laboratory limits on  colored particles like leptoquarks. Nevertheless, it is worth pointing out that the IceCube UHE neutrino data provides a new constraint on the leptoquark mass-coupling plane, as shown in Fig.~\ref{fig:LQ}.

\subsection{Collider Constraints} \label{sec:LQLHC}
\begin{figure}[t!]
    \centering
    $$
    \includegraphics[width=.45\textwidth]{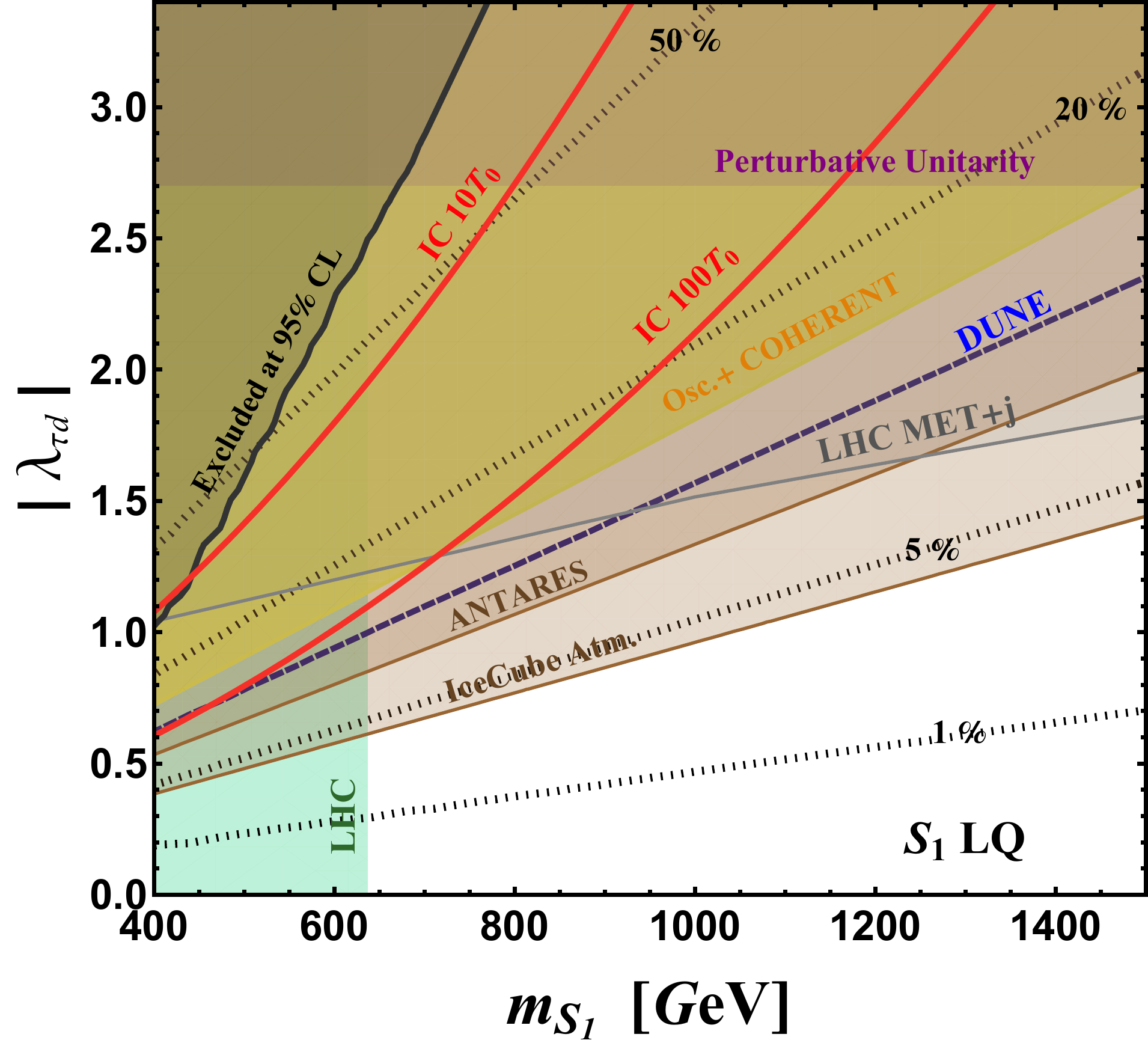} \hspace{0.2in}
     \includegraphics[width=.45\textwidth]{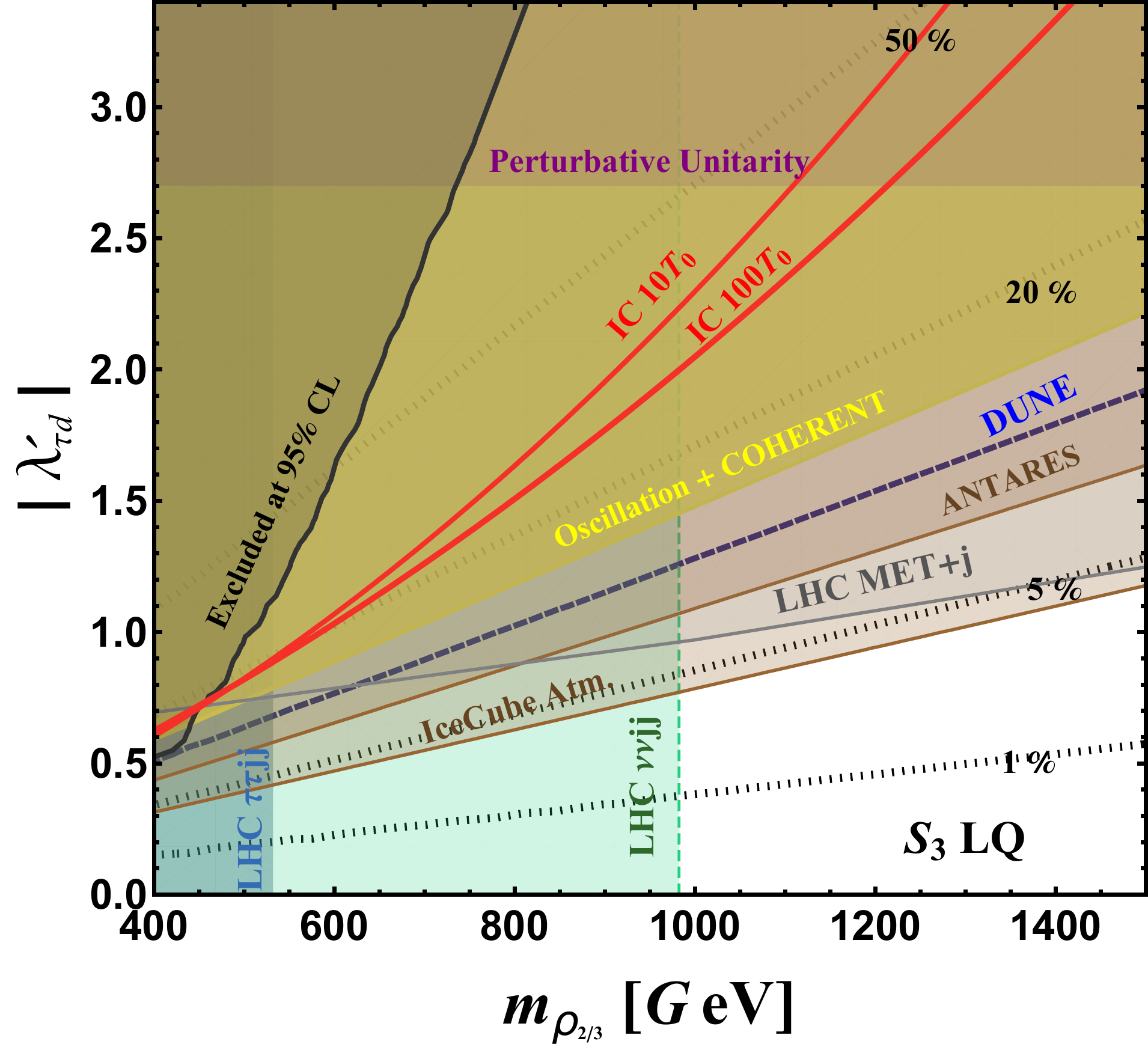}
     $$
       $$
    \includegraphics[width=.45\textwidth]{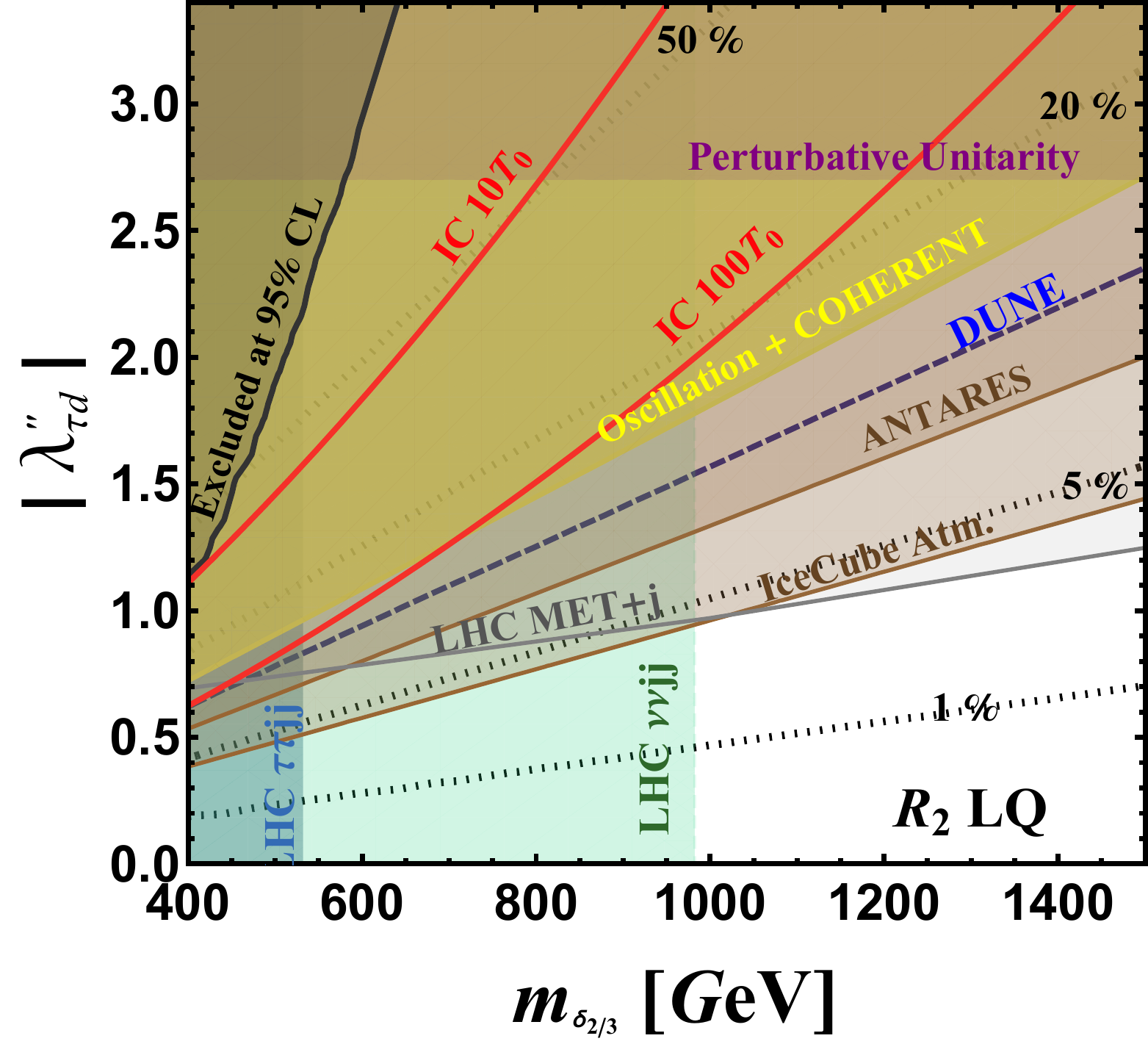} \hspace{0.2in}
     \includegraphics[width=.45\textwidth]{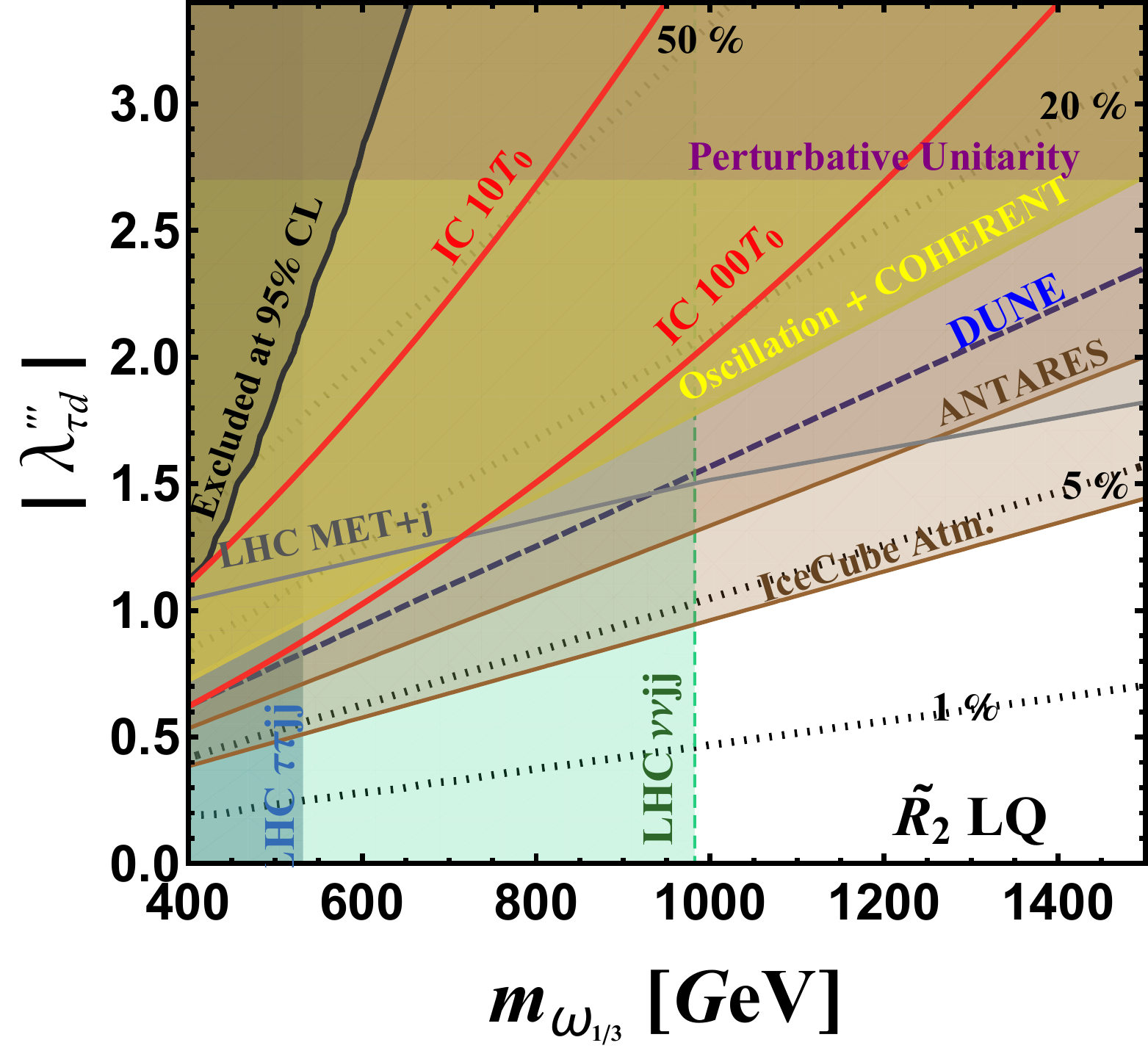}
     $$
    \caption{IceCube sensitivity contours (corresponding to one expected event in the resonance energy bins combined) for each of the scalar leptoquark  parameter space relevant for $\varepsilon_{\tau\tau}$ are shown by the thick red curves, for different exposure times (in terms of the current exposure $T_0=2653$ days). The black-shaded region is the current exclusion derived from a bin-by-bin analysis of the IceCube data (see Fig.~\ref{fig:LQspectrum}).  
  The predictions for  $\varepsilon_{\tau\tau}$ are shown by the thin dotted contours. The shaded regions are excluded: green-shaded by LHC $\nu j \nu j$~\cite{CMS:2018qqq}, blue-shaded by LHC $\tau j \tau j$~\cite{ATLAS:2019qpq}, grey-shaded by LHC monojet plus MET~\cite{Aaboud:2017phn}, purple-shaded by perturbative unitarity~\cite{DiLuzio:2017chi}, yellow-shaded by global fit to neutrino oscillation plus COHERENT data~\cite{Coloma:2019mbs}, and light (dark) brown by IceCube~\cite{IceCubeCollaboration:2021euf} (ANTARES~\cite{Albert:2021sfx}) atmospheric neutrino data. For more details on these exclusion regions, see Ref.~\cite{Babu:2019mfe}. For comparison, we also show the future DUNE sensitivity in blue dashed line~\cite{Chatterjee:2021wac}. 
  }
  \label{fig:LQ}
\end{figure}

In this section, we analyze the collider implications of the scalar leptoquarks  $S_1, R_2, \tilde{R}_2, S_3$. At the LHC, leptoquarks are dominantly produced in pairs via gluon fusion process (and sub-dominantly via Drell-Yan process), and the pair-production rate is uniquely determined by the leptoquark masses. 
There could be single production of leptoquarks in association with leptons as well; however, the single production rate is dependent on both the Yukawa couplings as well as the leptoquark mass. It is worth mentioning that the LHC limits from single-production of leptoquarks are not so stringent compared to the pair-production limits \cite{ATLAS:2020dsk, ATLAS:2019qpq} unless the Yukawa couplings to first and second-generation quarks and leptons are
$\gtrsim {\cal O}(1)$. A detailed discussion and comparison between the single- and pair-production limits can be found in Refs.~\cite{Babu:2019mfe,Buonocore:2020erb}. Since we are not considering any large Yukawa couplings via which leptoquarks talk to quarks {\it and} leptons from first or second generation, the limits from single-production are not applicable here. In order to allow for the lowest possible leptoquark mass which will give the maximum signal at IceCube, we consider an optimized setup where the leptoquarks communicate with neutrinos of tau flavor and first-generation quarks only. All other entries in the Yukawa coupling matrix $\lambda_{ij}$ are kept small. This scenario can be made consistent with neutrino oscillation data in models of radiative neutrino masses~\cite{Babu:2019mfe}. Thus, in our setup, the dominant bounds on leptoquark masses will come from $\nu j \nu j$ and $\tau j \tau j $ searches at the LHC. There are dedicated searches for leptoquarks in the $\nu j \nu j$ channel~\cite{CMS:2018qqq} which we will adopt in our analysis to recast the pair-production bound. However, there are no specific searches for the $\tau j \tau j$ final state. For constraints on $\lambda_{\tau d}$, we recast the $\tau^+ \tau^- b\bar{b}$ search limits~\cite{ATLAS:2019qpq} taking into account the $b$-jet misidentification as light jets, with an average rate of $1.5\%$ (for a $b$-tagging efficiency of 70\%)~\cite{Chatrchyan:2012jua}. Our results are shown in Fig.~\ref{fig:LQ}, where the green and blue-shaded regions are excluded by $\nu j \nu j$ and $\tau j \tau j$ channels respectively in each leptoquark case.

Note that the limit from $\tau j \tau j$ searches is less severe than the $\nu j \nu j $ search limit. The $S_1$ leptoquark, being of single component, decays to $\tau j$ and $\nu j$ with 50$\%$ branching ratio each. As we can see from Fig.~\ref{fig:LQ} (top left panel)  the strongest bound on $S_1$ leptoquark mass therefore comes from the $\nu j \nu j$ search and excludes LQ masses below $\sim$ 630 GeV, independent of the Yukawa coupling. For the $R_2$ leptoquark, there are two different components ($\delta^{2/3}$ and $\delta^{5/3}$), out of which the former is responsible for generating NSI   $\varepsilon_{\tau \tau}$. At the LHC, the  pair-production of the leptoquark component $\delta^{2/3}$ will lead to $\nu j \nu j$ signature, whereas its $\delta^{5/3}$ counterpart will lead to $\tau j \tau j$ signature. In principle, we could make a judiciary choice that $\delta^{2/3}$ and $\delta^{5/3}$ components are split by a mass $\Delta m=|m_{\delta^{2/3}}-m_{\delta^{5/3}}|>m_W$ such that the $\delta^{2/3}$ component can decay back to $\delta^{5/3}$ in association with a $W$ boson, thus suppressing its branching ratio to $\nu j$ and hence the LHC limit~\cite{Babu:2019mfe}. However, for any non-degenerate $SU(2)$ doublet, the maximum mass splitting $\Delta m$ allowed by the electroweak $\rho$-parameter constraint at 90\% CL   is~\cite{ParticleDataGroup:2020ssz} 
\begin{align}
    \Delta m < \sqrt{\frac{3}{C}}\ (48~{\textrm{GeV}}) \, , 
\end{align}
where $C=1\ (3)$ for color singlets (triplets). This prevents the on-shell two-body decay of one leptoquark component to a lighter one, accompanied by a $W$-boson. The three-body decay with an off-shell $W$-boson is still possible, but its branching ratio with respect to the dominant $\nu j$ branching is very suppressed (to less than percent level).  Therefore, we will just use the LHC limits on each of the leptoquark components $\delta^{2/3}$ and $\delta^{5/3}$ assuming 100\% branching ratio into the $\nu j$ and $\tau j$ channels, respectively, with the corresponding mass bounds of 630 GeV and 532 GeV, respectively. The same limits are also applicable for the $\tilde{R}_2$ and $S_3$ leptoquarks. 

In addition to the pair- and single-production limits from LHC searches, there will be a direct limit on the NSI strength $\varepsilon_{\tau \tau}$ from the  mono-jet plus missing transverse energy (MET) signature at the LHC. For leptoquarks, this arises via a $t-$ channel leptoquark exchange: $pp \to \nu \bar{\nu} j$, where the gluon jet originates from the initial state radiation off one of the quark legs. The corresponding constraint in the leptoquark mass and coupling plane has not been reported before; so we undertake this analysis here. We simulate the $pp\rightarrow \nu\bar{\nu}j$ signal events with {\sc MadGraph5aMC@NLO}~\cite{Alwall:2014hca} event generator, then analyze the hadronization and parton shower effects with {\sc Pythia8}~\cite{Sjostrand:2007gs}, and detector effects with {\sc Delphes3}~\cite{deFavereau:2013fsa}. There is a dedicated LHC search for dark matter in the monojet+MET channel~\cite{Aaboud:2017phn}. We follow this analysis and implement similar acceptance criteria: (a)  we define jets with the anti-$k_t$ jet algorithm and radius parameter $R=0.4$, pseudo-rapidity  $|\eta|<2.8$ and $p_{Tj}>30$~GeV  via {\sc FastJet}~\cite{Cacciari:2011ma}; (b) we veto the events with identified electrons with $p_T>20$~GeV or muons $p_T>10$~GeV in the final state; (c)  in order to reduce the $W$+jets and $Z$+jets backgrounds, we select the events with ${\rm MET}>250$~GeV  recoiling against a leading jet with $p_{Tj1}>250$~GeV, $|\eta_{j1}|<2.4$, and azimuthal separation $\Delta\phi(j_1,\vec{p}_{T,{\rm miss}})>0.4$; and (d) we veto the events with more than four jets. We find that the limits from our monojet +MET study are the most stringent LHC limits on NSI for large leptoquark mass regimes. In Fig.~\ref{fig:LQ}, this bound is depicted by the grey-shaded regions. As we can see, the limits from different collider searches on leptoquarks preclude the possibility of finding a leptoquark resonance at IceCube or its future extension like IceCube-Gen2.

\section{New Resonances Induced by Pseudo-Dirac Neutrinos} \label{sec:pseudo}

\begin{figure*}[t!]
  \centering
 \includegraphics[width=0.49\textwidth]{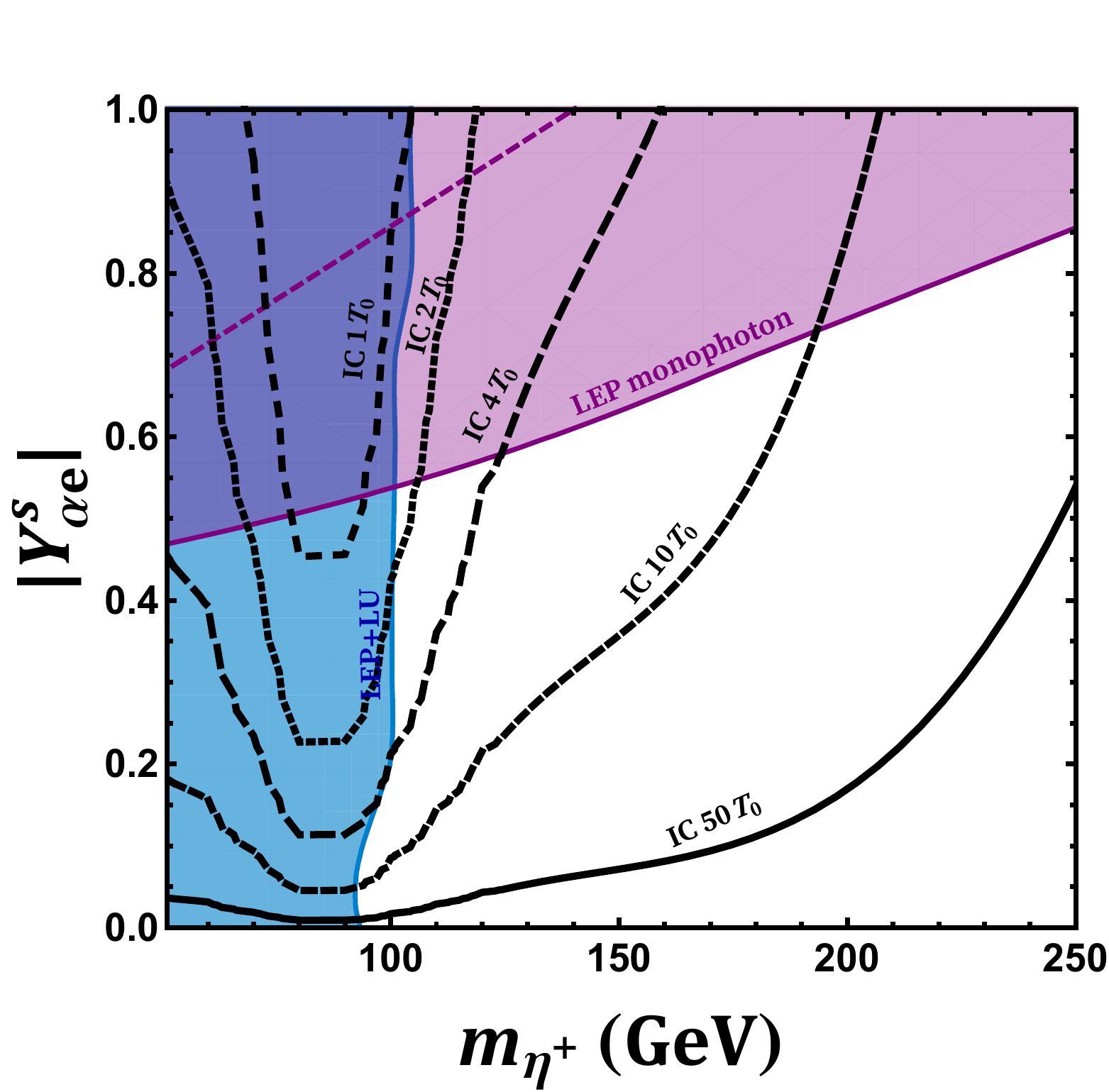}
  \caption{IceCube sensitivity contours (thick black curves, corresponding to one expected event in the resonance energy bins combined with different exposures) for the pseudo-Dirac neutrino case where the resonance is induced by an incoming sterile neutrino, and therefore, some of the NSI constraints shown in Fig.~\ref{fig:NSI} are no longer applicable. Other labels are the same as in Fig.~\ref{fig:NSI}. 
  }
  \label{fig:NSIPD}
\end{figure*}

In another interesting class of neutrino mass models, neutrinos are pseudo-Dirac particles~\cite{Wolfenstein:1981kw, Petcov:1982ya}. Depending on the mass splitting and mixing between the left-handed and right-handed (or sterile) neutrinos, one could distinguish the pseudo-Dirac case from purely Majorana or purely Dirac cases. Here we explore the possibility of inducing new resonances at IceCube by pseudo-Dirac neutrinos. The important question is: How do we produce them in UHE neutrinos? As has been pointed out in Refs.~\cite{Kobayashi:2000md, Beacom:2003eu, Keranen:2003xd}, for a tiny mass splitting $\delta m^2\lesssim 10^{-12}~{\rm eV^2}$ and maximal mixing, the active neutrino produced in the standard astrophysical sources can just oscillate into the sterile one as it reaches the earth. This would also reduce the active neutrino flux by half, but this is still consistent with the observed astrophysical neutrino flux, given the large uncertainty on the flux normalization and source properties. The flavor ratio measurements at IceCube can be sensitive to mass splittings as small as $10^{-19}~{\rm eV}^2$. Note that such tiny mass splittings have no effect on the solar~\cite{deGouvea:2009fp}  neutrino flux, and are also consistent with the BBN bounds~\cite{Barbieri:1989ti, Enqvist:1990ek}. According to Ref.~\cite{Martinez-Soler:2021unz}, there is in fact a mild preference for a non-zero mass splitting of $6\times 10^{-20}~{\rm eV}^2$ in the SN1987A data. 

As an example, let us consider the interaction Lagrangian for the sterile component of a pseudo-Dirac neutrino with an $SU(2)_L$-singlet charged scalar:
\begin{align}
    {\cal L} \supset Y^s_{\alpha \beta} \overline{\nu_{s \alpha}^c} \eta^+ \ell_{\beta R} +{\rm H.c.}
\end{align}
For the flavor index $\beta=e$, this could lead to an $\eta^+$ resonance at IceCube. 
As for the resonance feature, the event spectrum analysis for the pseudo-Dirac case will be exactly the same as the Majorana case discussed above, with the appropriate flux normalization. One of the main differences will be in the nature of mediator in Fig.~\ref{fig:feyn_spectra}. In particular, the role of $SU(2)_L$-singlet and doublet mediators will be interchanged for incoming sterile neutrinos, which are $SU(2)_L$-singlets as opposed to the $SU(2)_L$-doublet active neutrinos. Another major difference lies in the laboratory constraints, because none of the low-energy constraints on NSIs in Figs.~\ref{fig:NSI} and \ref{fig:LQ} will be applicable anymore for the sterile interactions. However, the collider constraints from LEP and LHC remain unchanged. Thus, we open up more parameter space for the Zee-type scalar resonances in Fig.~\ref{fig:NSI}, as shown explicitly in Fig.~\ref{fig:NSIPD}, but not for the leptoquark resonances in Fig.~\ref{fig:LQ}.        

\section{Conclusion} \label{sec:con}
We have discussed the possibility of new Glashow-like resonances in neutrino telescopes like IceCube, induced by scalar mediators of radiative neutrino mass mechanism. The same interactions that lead to the new signatures in UHE neutrino interactions at IceCube also give rise to observable non-standard interactions of neutrinos with matter, so that the UHE neutrinos provide a new complementary probe of NSI. We have provided explicit realizations of this idea taking the popular Zee model of radiative neutrino mass with extra scalars and its colored variants with leptoquarks. We find that the Zee-scalar resonances can be probed at IceCube and its future extensions like IceCube-Gen 2. The results are not so promising for  the scalar leptoquark resonances because the parameter space leading such resonances observable in foreseeable IceCube data is already ruled out by the LHC constraints. In this context, we derived new constraints on the leptoquark parameter space from LHC monojet plus MET search. We also discussed the pseudo-Dirac neutrino case where the IceCube detection prospects could be enhanced. To conclude, UHE neutrinos at neutrino telescopes provide a new probe of relatively light scalars, complementary to other low-energy and collider searches.

\section*{Acknowledgments}

We thank Yicong Sui for earlier collaboration, and Werner Rodejohann for useful discussions. The work of KSB is supported by the US Department of Energy Grant No.~DE-SC 0016013. The work of BD is supported in part by the  US  Department  of  Energy  under Grant No.~DE-SC0017987. 

{\footnotesize
\bibliographystyle{utphys}
\bibliography{ref}
}
\end{document}